\documentclass[conference]{IEEEtran}
\usepackage{cite} 
\usepackage{amsmath,amssymb,amsfonts} 
\usepackage{graphicx} 
\usepackage{textcomp} 
\usepackage{xcolor} 
\usepackage[utf8]{inputenc}
\usepackage{amsmath}
\usepackage{multirow}
\usepackage{array}

\usepackage{cite}
\usepackage{amsmath,amssymb,amsfonts}
\usepackage{graphicx}
\usepackage{textcomp}
\usepackage{graphicx} 
\usepackage{subcaption}
\def\BibTeX{{\rm B\kern-.05em{\sc i\kern-.025em b}\kern-.08em
    T\kern-.1667em\lower.7ex\hbox{E}\kern-.125emX}}

\usepackage{algorithm} 
\usepackage{algpseudocode} 

\begin{document}

\title{Breaking Barriers in Health Monitoring: Multi-Scenario Vital Sign Detection Using Mm-Wave MIMO FMCW Radar}

\author{\IEEEauthorblockN{Ehsan Sadeghi}
\IEEEauthorblockA{\textit{EEMCS Faculty} \\
\textit{University of Twente}\\
Enschede, The Netherlands \\
e.sadeghi@utwente.nl}
\and
\IEEEauthorblockN{Paul Havinga}
\IEEEauthorblockA{\textit{EEMCS Faculty} \\
\textit{University of Twente}\\
Enschede, The Netherlands \\
p.j.m.havinga@utwente.nl}
}

\maketitle

\begin{abstract}
This paper explores the deployment of mm-wave Frequency Modulated Continuous Wave (FMCW) radar for vital sign detection across multiple scenarios. We focus on overcoming the limitations of traditional sensing methods by enhancing signal processing techniques to capture subtle physiological changes effectively. Our study introduces novel adaptations of the Prony and MUSIC algorithms tailored for real-time heart and respiration rate monitoring, significantly advancing the accuracy and reliability of non-contact vital sign monitoring using radar technologies. Notably, these algorithms demonstrate a robust ability to suppress noise and harmonic interference. For instance, the mean absolute errors (MAE) for MUSIC and Prony in heart rate detection are 1.8 and 0.81, respectively, while for respiration rate, the MAEs are 1.01 and 0.8, respectively. These results underscore the potential of FMCW radar as a reliable, non-invasive solution for continuous vital sign monitoring in healthcare settings, particularly in clinical and emergency scenarios where traditional contact-based monitoring is impractical.

\end{abstract}


\section{Introduction}
\label{sec:introduction}

The continuous and accurate monitoring of vital signs such as heart rate and respiratory rate is indispensable across a range of settings including clinical environments, remote patient care, and elderly care \cite{ref00}. These indicators are pivotal for diagnosing conditions, managing chronic illnesses, and ensuring timely medical responses in emergency situations. Traditionally, vital signs have been monitored using direct contact methods such as electrocardiography (ECG) and photoplethysmography (PPG) \cite{intro-1,intro-1-2}. However, these methods can be cumbersome, invasive, and are often unsuitable for long-term or non-intrusive monitoring \cite{intro-2}.

Given the limitations of physical contact methods, there has been growing interest in non-contact sensing technologies. Techniques involving cameras, thermal cameras, and other optical sensors have been explored in both humans and animals, but they frequently fall short in low visibility conditions and raise privacy concerns. Moreover, their dependence on environmental factors significantly restricts their reliability and effectiveness \cite{ref01, ref02}.

Recently, radio frequency-based monitoring using radar 
has emerged as a robust alternative for remote health monitoring applications. Radar systems are highly effective in diverse environments and can operate under conditions where other non-contact methods fail.
Recently, radars have been utilized for a wide range of applications like vital sign detection and activity recognition in both humans and animals\cite{MySurvey, RayPet, AR-ex}.
In the context of vital sign detection, radar targets cardiopulmonary chest displacements to extract movements related to the respiratory and cardiovascular systems, revealing heart rate and respiration rate. These movements are quite subtle, with chest displacement due to respiration ranging from 4 mm to 12 mm and due to cardiac activities from 0.2 mm to 0.5 mm \cite{intro-3}.
Various types of radar, such as ultra-wideband (UWB) radar \cite{intro-4}, stepped frequency continuous wave radar \cite{intro-5}, and continuous wave (CW) radar \cite{intro-6}, have been adopted for this purpose. However, Frequency Modulated Continuous Wave (FMCW) radars—particularly in the mm-wave band—offer superior resolution, accuracy, and stability, making them well-suited to detect these minute physiological movements.

While remote monitoring of vital signs using FMCW radar shows promise, several challenges remain. Although FMCW radar accurately estimates respiration rate, heart rate estimation is notably more difficult. This is primarily due to the dominance of respiratory signals over the more subtle cardiac signals, with higher harmonics of the respiration rate occasionally overlapping the frequencies associated with the heart rate, thereby complicating heart rate estimation. Furthermore, the low amplitude of the heart rate signal makes it susceptible to various interferences such as reflections from stationary objects, transmitter-receiver non-linearity, and impulsive noise.

Various studies have explored and developed methods for heart rate and respiration rate extraction to overcome challenges in non-contact vital sign monitoring. Giordano et al. utilized both the Fast Fourier Transform (FFT) and Short Time Fourier Transform (STFT) for heart rate estimation in 16 individuals positioned 45 to 55 cm away from the radar, achieving an average relative error of 7.4\% \cite{myref-5}. Although FFT-based methods have been prevalent, they often fall short in addressing the complexities of heart rate estimation, prompting further research into more robust techniques like the Kalman filter and Rife algorithm \cite{myref-4, myref-6}.
Ji et al. introduced a method named SHB, incorporating Kalman filtering and peak detection, which demonstrated an ability to achieve an average root mean square error (RMSE) of 3.024 for five subjects tested at a distance of 50 cm from the radar \cite{myref-4}. Similarly, Chen et al. combined the Kalman filter and Rife algorithm to attain an average RMSE of 4.6 BPM for a test group of 10 subjects situated 80 cm away from the radar \cite{myref-6}.

Despite their promise, several challenges persist. While respiration can typically be estimated accurately, heart rate detection is much more difficult. The cardiac signal is low in amplitude and often masked by higher harmonics of respiration. Moreover, the heart rate signal is susceptible to interference from static reflections, non-linearities in the radar hardware, and impulsive noise. As a result, many studies have attempted to refine the estimation process. Early approaches relied on spectral methods such as the Fast Fourier Transform (FFT) or Short-Time Fourier Transform (STFT) \cite{myref-5}, later augmented by filters and peak detection methods such as Kalman filtering and the Rife algorithm \cite{myref-4,myref-6}. More recent works have turned to decomposition-based approaches such as Empirical Mode Decomposition (EMD), Ensemble EMD (EEMD), and Independent Component Analysis (ICA) \cite{myref-14,myref-9,myref-10,myref-8}. While these methods improve accuracy, they often face challenges such as mode mixing, parameter sensitivity, or high computational cost, which hinder their use in practical real-time monitoring.

To address these limitations, this work introduces a new class of methods for vital sign extraction from FMCW radar. Specifically, we present the first adaptation and reformulation of the Multiple Signal Classification (MUSIC) and Prony algorithms for cardiopulmonary monitoring. Originally developed for high-resolution spectral estimation in array processing and signal decomposition, these algorithms are here tailored to the unique structure of radar-derived chest displacement signals, enabling accurate extraction of heart and respiration rates even under challenging conditions. To our knowledge, this represents the first application of MUSIC and Prony methods in radar-based vital sign detection.

\textbf{The main contributions of this paper are as follows:}
\begin{itemize}
    \item \textbf{Novel extraction methods:} We reformulate the MUSIC and Prony algorithms for application in FMCW radar vital sign monitoring. These adaptations allow the algorithms to separate cardiac signals from respiration harmonics and noise with significantly improved resolution compared to FFT- or decomposition-based techniques.
    \item \textbf{Signal-level and preprocessing enhancements:} We improve signal-to-noise ratio (SNR) through adaptive range-bin selection, average chirp integration, and MIMO channel fusion, alongside clutter removal, DC offset correction, impulsive noise suppression, and phase enhancement.
    \item \textbf{Scenario validation:} We extensively test the proposed algorithms across multiple scenarios, including varying distances, orientations, and angles, as well as special cases such as post-exercise elevated cardiopulmonary rates, asthma (high respiration rates), and meditation (low respiration rates).
    \item \textbf{Comparison with existing methods:} We benchmark our approaches against FFT and coarse-to-fine estimation strategies, demonstrating clear performance gains. Discussion also includes positioning relative to decomposition-based methods (e.g., EEMD, ICA).
    \item \textbf{Real-time feasibility:} We discuss the influence of observation window length and sliding-window processing on achieving real-time monitoring in practice. 
     \item \textbf{Enhanced FFT and coarse-to-fine estimation:} Alongside MUSIC and Prony, we strengthen baseline methods by improving FFT estimation and implementing a coarse-to-fine strategy, enabling more robust detection across varying conditions. 
\end{itemize}

The remainder of this paper is organized as follows: Section II reviews FMCW radar fundamentals for vital sign estimation. Section III describes the proposed approach. Section V introduces the experimental setup and dataset. Section VI presents results across scenarios, while Section VII provides discussion. Section VIII concludes the paper and outlines future research directions.

\section{Fundamentals of FMCW Radar for Vital Sign Detection}

\begin{figure}[t]
    
\includegraphics[width=9 cm]{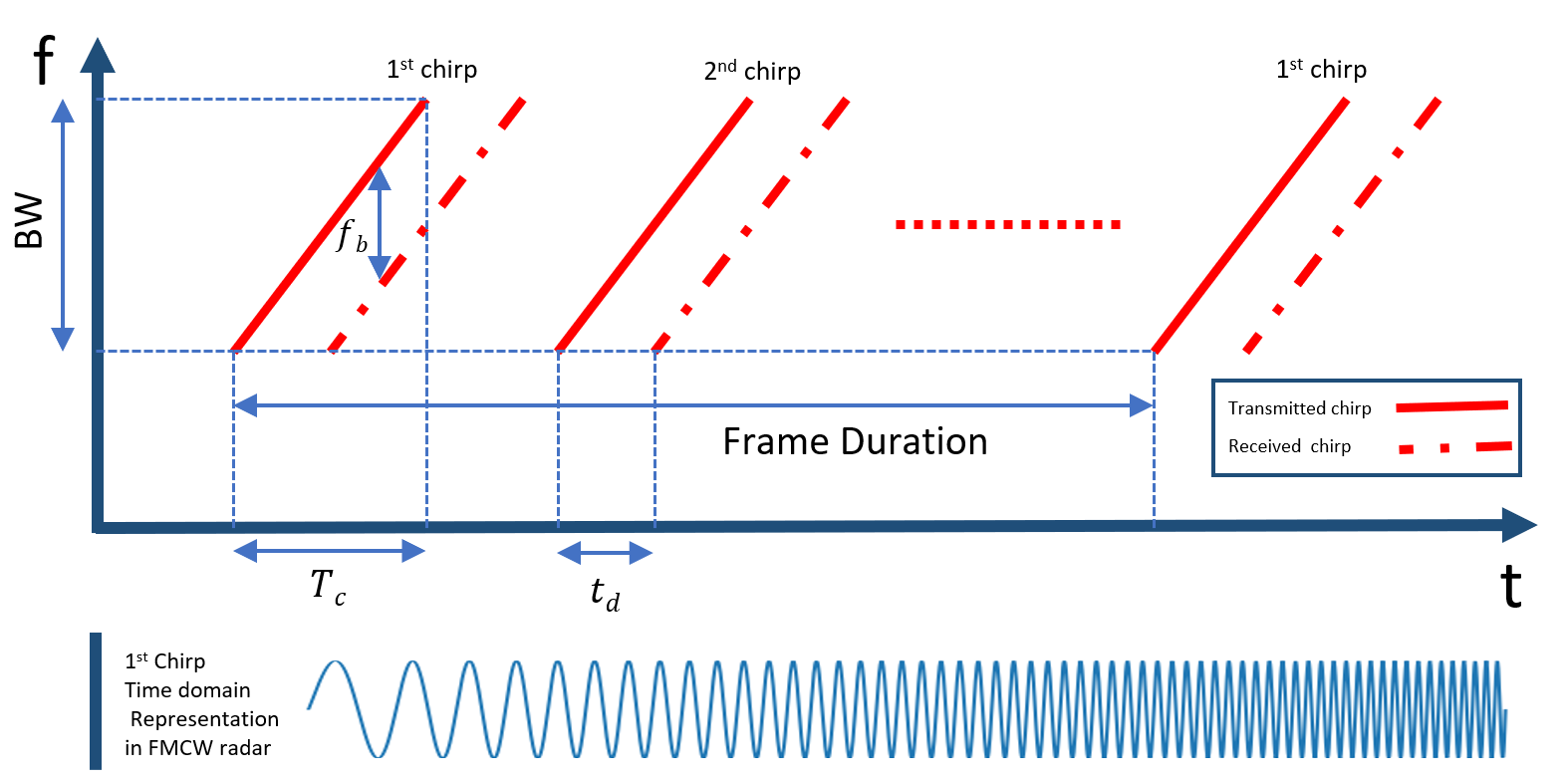}

\caption{Schematic of FMCW radar's transmitted and received signals.}
\label{radar-one}
\end{figure}

Millimeter-wave Frequency-Modulated Continuous-Wave (mm-wave FMCW) radar offers exceptional resolution and privacy in detecting human vital signs. Due to its ability to operate over a wide bandwidth, it provides high range resolution while the mm-wave domain ensures privacy by not revealing identifiable features.
FMCW radar determines the range of a target by sending out a chirp with a linearly increasing frequency. The difference in frequency between the transmitted and received signals is proportional to the target's distance, allowing the detection of vital signs with high precision. This mechanism offers an advantage over pulsed radar by being more power-efficient and offering simultaneous range and velocity measurements.
Fig. \ref{radar-one}, illustrates the schematic of transmitted and received chirps within a frame period in FMCW radar.

The FMCW radar system typically consists of a Low Noise Amplifier (LNA), Power Amplifier (PA), mixer, and Analog-to-Digital Converter (ADC). In the mixer, transmitted and received signals combine to produce the Intermediate Frequency (IF) signal, which is then sampled by the ADC.
Due to I/Q demodulation, the ADC samples are in complex format.
The transmitted signal \( S_{TX}(t) \) in FMCW radar can be formulated as:

\begin{equation}
S_{TX}(t) = A_{T} \cos(2\pi f_c t + \frac{\pi BW}{T_c}t^2 + \phi(t)), \,0<t<T_c
\end{equation}

with \( A_{T} \), \( f_c \), \( BW \), and \( T_c \) representing the amplitude, starting frequency, sweeping bandwidth, and chirp duration, respectively.
As a result, the received signal \( S_{RX}(t) \), containing information about the target's movement, can be expressed as:

\begin{equation}
S_{RX}(t) = A_{R} \cos(2\pi f_c (t-t_d) + \frac{\pi BW}{T_c}(t-t_d)^2 + \phi(t-t_d)),
\end{equation}
where \( t_d \) is the time delay corresponding to the round-trip time of the radar signal. Assuming the target's position is represented by \( R_0 \), \( t_d \) can be expressed as:

\begin{equation}
    t_d= \frac{2 \times R_0}{C},
\end{equation}
in which \( C \) is the speed of light.
I/Q demodulation is a crucial step in FMCW radar signal processing. This technique decodes the received signal into two orthogonal components, In-phase (I) and Quadrature (Q), to extract detailed phase information.
After processing through the mixer, the demodulated I/Q components of the IF signal, after low-pass filtering and approximations considering parameter ranges, can be expressed as:

\begin{equation}
S_{IF}(t) = A_{IF} \exp(j(2\pi f_b t + \phi_b(t) )),
\end{equation}
where \( f_b \) is the beat frequency and \( \phi_b(t) \) represents the phase change related to the target's tiny movements (chest displacement).
It is possible to show \( f_b \) and \( \phi_b(t) \) are equivalent to: 
\begin{equation}
    f_b= \frac{2B R_0}{C T_c},
    \label{fb}
\end{equation}
\begin{equation}
    \phi_b (t)=\frac{4\pi (r(t)+R_0)}{\lambda},
    \label{phase formula}
\end{equation}
where \( \lambda \) is the wavelength of the signal.
For vital sign detection, target displacements are modeled by \( R_0 + r(t) \). Considering the target remains relatively stationary, \( R_0 \) indicates the distance of the target from the radar, and \( r(t) \) represents the cardiopulmonary chest displacement of the target.
\( r(t) \) reflects both static and dynamic components due to respiratory and cardiac motion. This movement modulates the phase of the IF signal as expressed in Equation (\ref{phase formula}). The phase changes in the IF signal thus serve as direct indicators of vital signs, enabling the extraction of respiration and heartbeat signals from the phase variations of the reflected radar waves.

\begin{figure*}[ht]
\centering
\includegraphics[width=\textwidth]{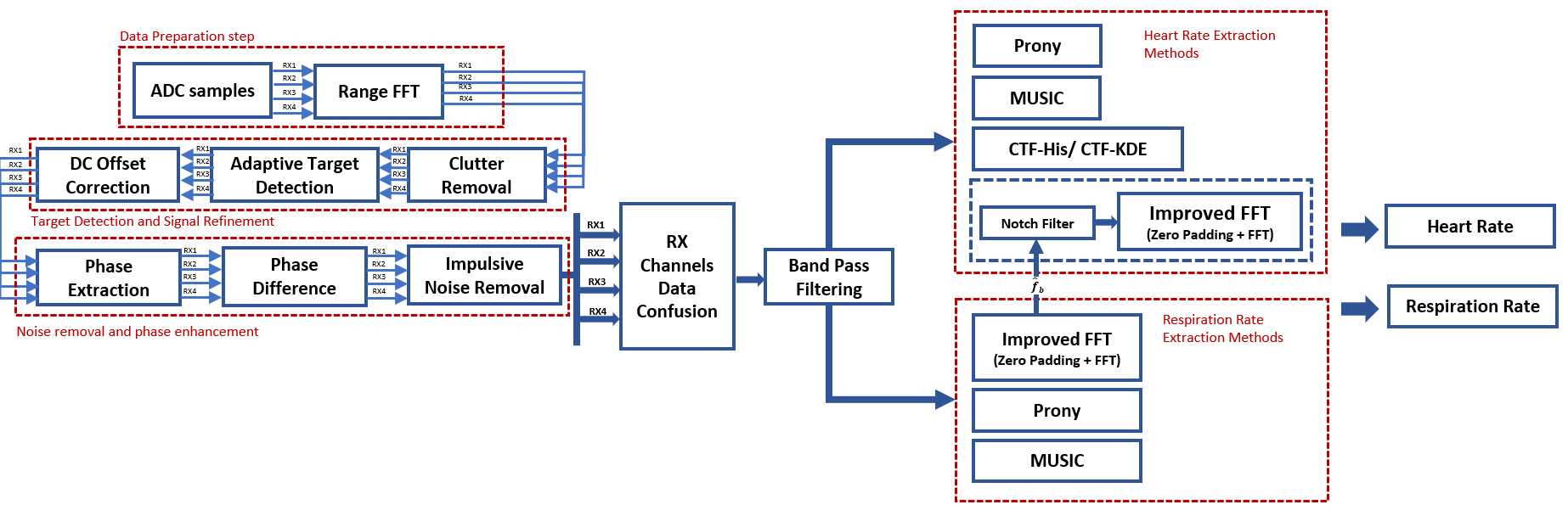} 
\caption{Overview of the Proposed Signal Processing Chain.}
\label{overview}
\end{figure*}

\section{Proposed approach}

The proposed signal processing chain is shown in Fig. \ref{overview}. The chain encompasses key steps such as data preparation, target detection and signal refinement, RX channel data processing, bandpass filtering, and heart and respiration rate extraction methods. After collecting the I/Q ADC samples of the IF signal, a range map is formed to help pinpoint the location of the target within the scene. Subsequently, the phase corresponding to the target is calculated, which is indicative of cardiopulmonary chest displacement. Several extensive signal processing steps are then required to refine the phase signal.
In the context of MIMO configurations, we process data from different receiver channels, integrating them before feeding the output signal through a bandpass filter. This filter is designed to isolate heart and respiration rate signals based on their distinct frequency ranges. Finally, we employ five and three different methods, including the Prony and MUSIC algorithms, to estimate heart and respiration rates, respectively.
Each step will be detailed in the sections that follow.

\subsection{Data Preparation step}

\subsubsection{ADC Samples and Range FFT}
\begin{figure*}[ht]
\centering
\includegraphics[width=\textwidth]{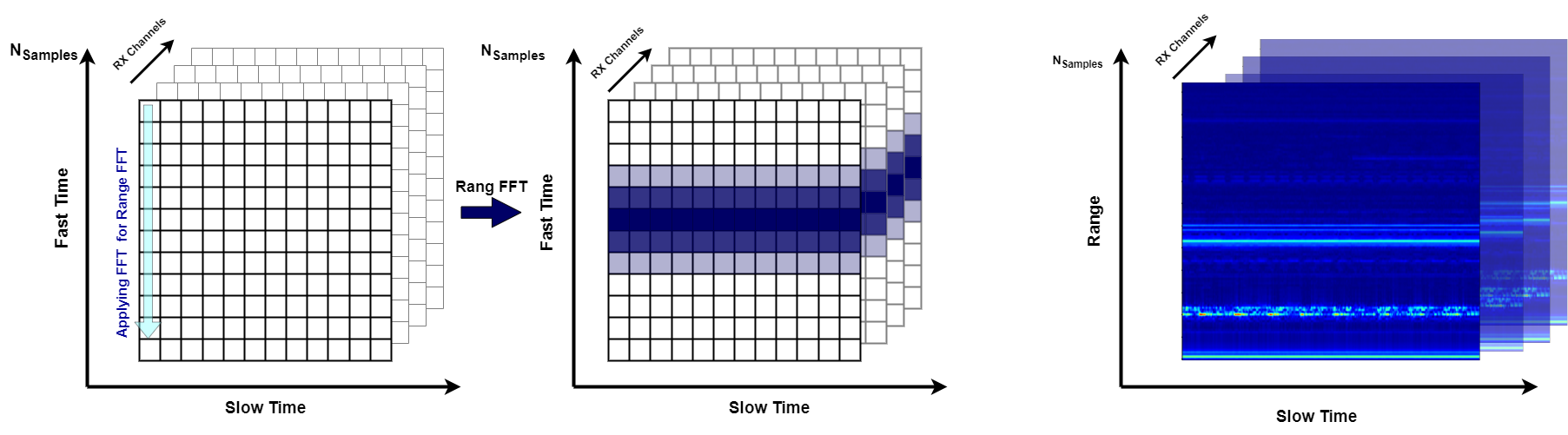} 
\caption{Overview of the Proposed Signal Processing Chain.}
\label{Range FFT}
\end{figure*}

The received I/Q samples from ADCs are organized into data cubes as illustrated in Fig. \ref{Range FFT}. In each column, $N_{\text{Samples}}$, which represent the ADC samples from each chirp, are stacked. Along the slow time axis, ADC samples from different chirps and different frames are combined. Additionally, in MIMO configurations, a similar data cube exists for each receiver channel.

Subsequently, a Range FFT is applied along each column (slow time axis) to analyze the spectrum of the beat signal and reveal range information. The outcomes of the Range FFT are depicted in Fig. \ref{Range FFT}. 
Each of the cubes in the horizontal axis represents range bins, which correspond to different beat frequencies ($f_b$) and distances. 
The range bin exhibiting the highest absolute value indicates the target's location. The relationship between range bin frequency and distance values can be expressed with the formula (\ref{fb}). Given the FFT resolution, the resolution of each range bin can be formulated as:

\begin{equation}
\text{Bin}_{\text{res}} = \frac{f_s T_c C}{2 BW N_{Samples}},
\end{equation}
where $f_s$ is the sampling rate of the radar. With our specified parameters, each range bin covers approximately 4 cm.

It can be shown that using multiple chirps and averaging them enhances signal-to-noise ratio (SNR) for vital sign detection, particularly heart and respiration rates. Unlike most studies focusing on the first chirp of each frame, we find that the brief 50-microsecond chirps maintain consistent vital sign indicators while noise varies. Due to the random characteristics of noise, averaging these chirps minimizes (or zeros out) noise and boosts relevant signals. This method enhances signal clarity and reliability for vital sign detection with FMCW radar, providing more accurate estimation in the following steps.

\subsection{Target Detection and Signal Refinement}

\subsubsection{Clutter Removal}
To enhance the accuracy of target location identification within a range map, it is crucial to eliminate the influence of static objects and their reflections, such as walls, furniture, and ceilings, present in the radar's surveillance area.
Since the chest displacement of the subject due to heart rate and respiration rate remains apparent across successive chirps, we can distinguish and eliminate static objects due to their invariable behavior in the range map while preserving vital signs information of the target.
Clutter removal is frequently employed in literature for this purpose, aiming to eliminate background static interference and suppress stationary clutter. 
The process involves the subtraction of the mean value from each range bin, a procedure that can be mathematically represented for each $k^{th}$ range bin as follows:




\begin{equation}
\bar{S}[k] = \frac{1}{ M} \sum^{ M}_{n=1} S[k,n]
\label{eq:averaging}
\end{equation}

\begin{equation}
S'[k,n] = S[k,n] - \bar{S}[k]
\label{eq:clutter_removal}
\end{equation}
In these equations, $S[k,n]$ denotes the FFT result for the $n^{th}$ observation time (corresponding to the $n^{th}$ frame) and the $k^{th}$ range bin (sample).
The symbol $\bar{S}[k]$ represents the average value calculated across each range bin over the set of received frames, whereas $S'[k,n]$ signifies the FFT values following the subtraction of the mean, thereby removing clutter. The constant M value denotes the total number of frames for each observation period.

\begin{figure}[h]
    \centering
    \begin{subfigure}[b]{0.45\textwidth}
        \includegraphics[width=\textwidth]{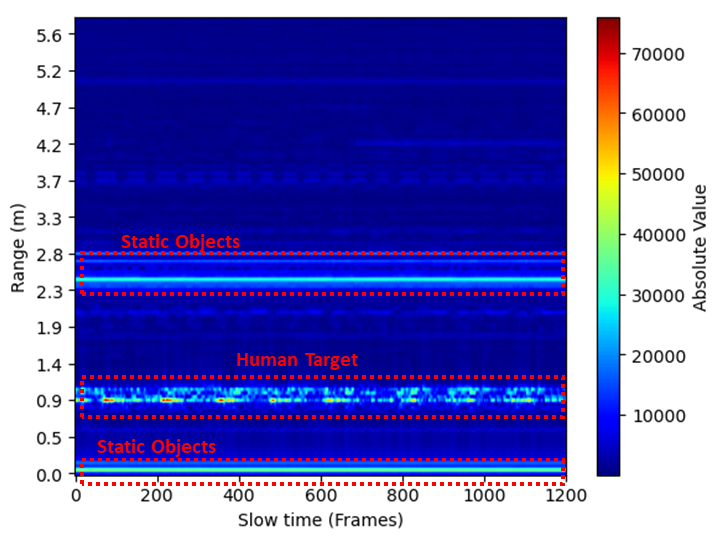}
        \caption{Before clutter removal}
        \label{fig:sub1}
    \end{subfigure}
    \hfill 
    \begin{subfigure}[b]{0.45\textwidth}
        \includegraphics[width=\textwidth]{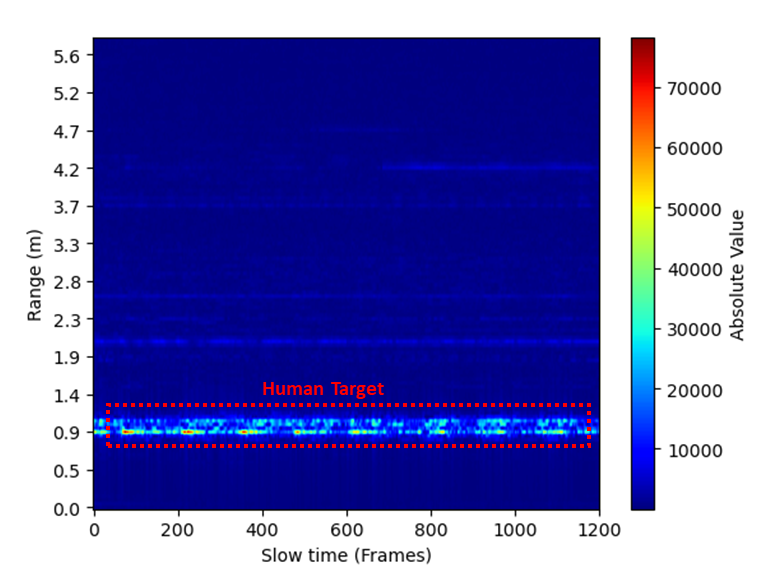}
        \caption{After clutter removal }
        \label{fig:sub2}
    \end{subfigure}
    \caption{Range map showing static and non-static reflections before and after clutter removal.
 }
    \label{Clutter1}
\end{figure}

\begin{figure}[h]
    \centering
    \begin{subfigure}[b]{0.45\textwidth}
        \includegraphics[width=\textwidth]{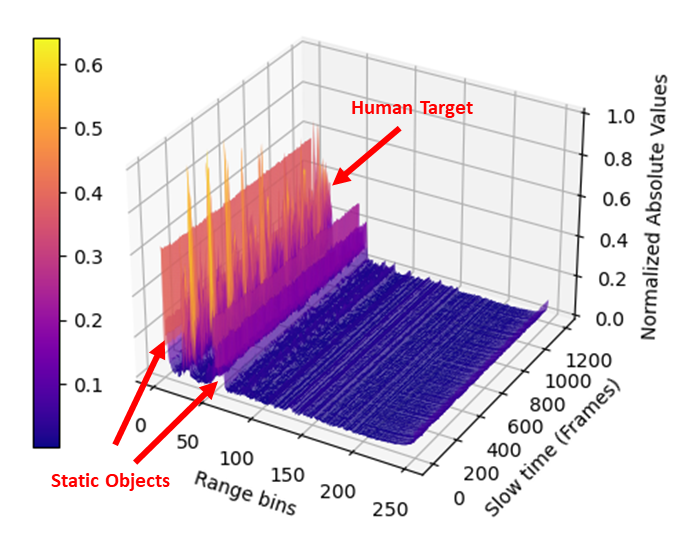}
        \caption{Before clutter removal}
        \label{fig:sub1}
    \end{subfigure}
    \hfill 
    \begin{subfigure}[b]{0.45\textwidth}
        \includegraphics[width=\textwidth]{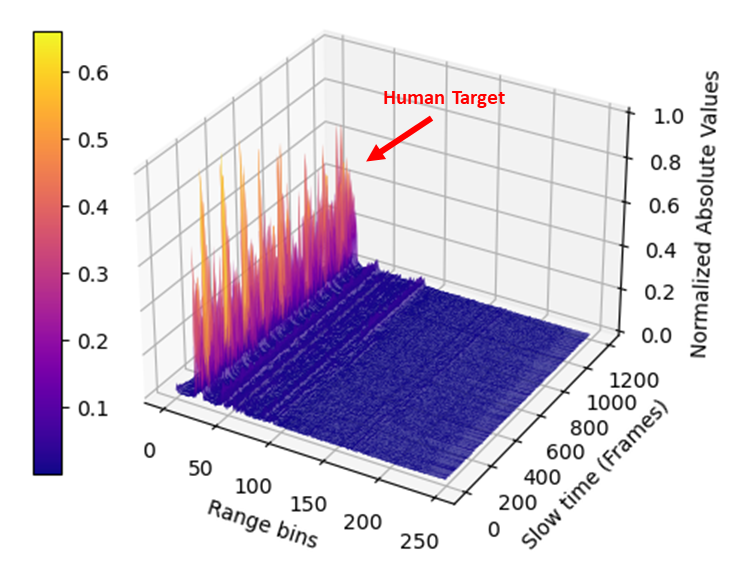}
        \caption{After clutter removal}
        \label{fig:sub2}
    \end{subfigure}
    \caption{Three-dimensional representation of range map showing static and non-static reflections before and after clutter removal.}
    \label{Clutter2}
\end{figure}


Following the removal of clutter and static reflections, the range map predominantly reflects non-static chest displacements driven by heart and respiration rates, which are expected to manifest as high absolute values. To clearly show and explain the impact of clutter removal, Figure \ref{Clutter1} and Figure \ref{Clutter2} display the range map (range slow time) before and after clutter removal. This significant reduction in static object reflections greatly assists in later steps aimed at accurately detecting the target in the range map and extracting its phase for further analysis.

\subsubsection{Adaptive Target Detection}

Considering the formula in (\ref{fb}), different range bins correspond to various objects within the radar scene. After implementing clutter removal to filter out reflections from static objects, the most potent signal in terms of power within the range map is typically that of the human target, primarily due to cardiopulmonary chest displacements. Consequently, the range bin exhibiting the highest absolute value in the FFT results, as depicted in Fig. \ref{Clutter1} (b), indicates the location of the target. As illustrated, the target is approximately 0.9 meters from the radar. By concentrating on this specific range bin and analyzing the IQ samples and phase data, we can proceed to calculate the phase changes associated with chest displacements of the target.

In adaptive target detection, we account for the possibility that phase and cardiopulmonary chest displacements may be distributed across multiple range bins. Therefore, we initially identify the range bin with the maximum value and then evaluate its two neighboring bins (considering the range of possible movements and chest displacements). If more than 20\% of the maximum values along the slow-time axis are found in these neighbors, rather than in a single range bin, we consider the average of these bins for phase calculation and subsequent processing steps. This approach ensures that we capture all relevant signal components and enhance the accuracy of our target detection.

\subsubsection{DC Offset Correction}



After identifying the target's location and its corresponding range bin(s), we proceed to extract the phase information of the target and its related Intermediate Frequency (IF) signal. However, the presence of any Direct Current (DC) values or offsets can result in inaccurate phase calculations.

Although the clutter removal process effectively reduces stationary clutter—a principal source of DC value offsets, as highlighted in \cite{AZ9}—other sources of DC offsets persist due to hardware issues. These include leakage between the Transmit (TX) and Receive (RX) antennas, local oscillator leakage, inadequate mixer isolation, non-linearity of the I and Q demodulator and mixer, as well as Radio Frequency (RF) cross-talk along the transmitter and receiver pathways \cite{AZ70}.

Such hardware-related issues can introduce additional DC components into the received signal, potentially compromising the accuracy of phase extraction.

Assuming $I(t)$ and  $Q(t)$ denote the In-phase and Quadrature-phase components of the demodulated signal, respectively, they are defined as follows:

\begin{equation}
\begin{split}
I(t) &= B_{I} \cos(2\pi f_{b} nT + \phi), \\
Q(t) &= B_{Q} \sin(2\pi f_{b} nT + \phi).
\end{split}
\end{equation}

where $f_{b}$ denotes the beat frequency,and $B_{I}$ as well as $B_{Q}$ represent the amplitudes of the signal in each I/Q channels.
However, the presence of DC components and offsets alters the received values to  $I^{DC}(t)$ and  $Q^{DC}(t)$, where $DC_{I}$ and $DC_{Q}$ indicate the DC components and offsets in the I and Q channels, respectively. Consequently, the estimated phase deviates from the actual phase associated with the target's heart and respiration-induced chest movements:

\begin{equation}
\begin{split}
I^{DC}(t) &= I(t) + DC_{I}, \\
Q^{DC}(t) &= Q(t) + DC_{Q}.
\end{split} \label{dcsq}
\end{equation}

\begin{equation}
\hat{\phi}(t)= \arctan\left( \frac{Q^{DC}(t)}{I^{DC}(t)} \right) \neq \phi(t)= \arctan\left( \frac{Q(t)}{I(t)} \right)
\end{equation}

To correctly estimate the phase of the target range bin we should find the values of $DC_{I}$ and $DC_{Q}$ which are I and Q offsets values or equivalently real and imaginary parts of offset.
 It can be shown that data points can form a circular constellation in the real and imaginary plane with the center of $DC_{I}$ and $DC_{Q}$. To estimate this centre and remove it (shifting the centre to zero) we take a similar approach to dynamic DC offset tracking done by the authors of \cite{myref-11}. By squaring both sides of \ref{dcsq}, summing these equations, and few modifications we can show that:

 \begin{equation}
\left( \frac{|I^{DC}(t) - DC_{I}|}{B_{I}} \right)^{2} + \left( \frac{|Q^{DC}(t) - DC_{Q}|}{B_{Q}} \right)^{2} = 1. 
\end{equation}

Assuming $B_{I}=B_{Q}=B_{R}$, where $B_{R}$ is the received amplitude. We move everything to left and call it $J(DC_{I},DC_{Q},B_{R})$:

\begin{equation}
|I^{DC}(t) - DC_{I}|^{2} + |Q^{DC}(t) - DC_{Q}| ^{2} - B^{2}_{R}=0. 
\end{equation}


To enhance system performance, our goal is to minimize the objective function $J(DC_{I},DC_{Q},B_{R})$, which is influenced by DC offset components and the amplitude of the received signal.
This minimization is crucial for the accurate determination of $DC_{I}$ and $DC_{Q}$ values, which further allows for the reconstruction of the $I(t)$ and  $Q(t)$ signal components. Ultimately, this process facilitates precise phase calculation, improving signal processing outcomes.
We employ the Gradient Descent (GD) algorithm for the optimization, leveraging its capability to efficiently find minimum values of the objective function. The mathematical formulation of our optimization problem and the gradients used in GD are as follows:

\begin{equation}
\min_{DC_{I}, DC_{Q}, B_{R}} J(DC_{I}, DC_{Q}, B_{R})
\end{equation}

\begin{equation}
J = \sum_n  \left(|I^{DC}[n] - DC_{I}|^{2} + |Q^{DC}[n] - DC_{Q}| ^{2}\right) - B^{2}_{R}
\end{equation}

\begin{equation}
\frac{\partial J}{\partial DC_{I}} = \sum_n 2 \left( DC_I - I^{DC}[n]\right)
\end{equation}

\begin{equation}
\frac{\partial J}{\partial DC_{Q}} = \sum_n 2 \left((DC_Q - Q^{DC}[n]\right)
\end{equation}

\begin{equation}
\frac{\partial J}{\partial B_{R}} = -2 B_{R}
\end{equation}


In our context, n represents the sample points of the target signal. Once the optimization process is completed, we obtain the values of $DC_{I}$ and $DC_{Q}$, which are then subtracted from the respective I and Q signal components.
This subtraction effectively eliminates the DC offset, thereby purifying the signal for further processing. Utilizing these corrected I and Q signals, we can accurately calculate the signal's phase.

\begin{figure}[h]
    
\includegraphics[width=9 cm]{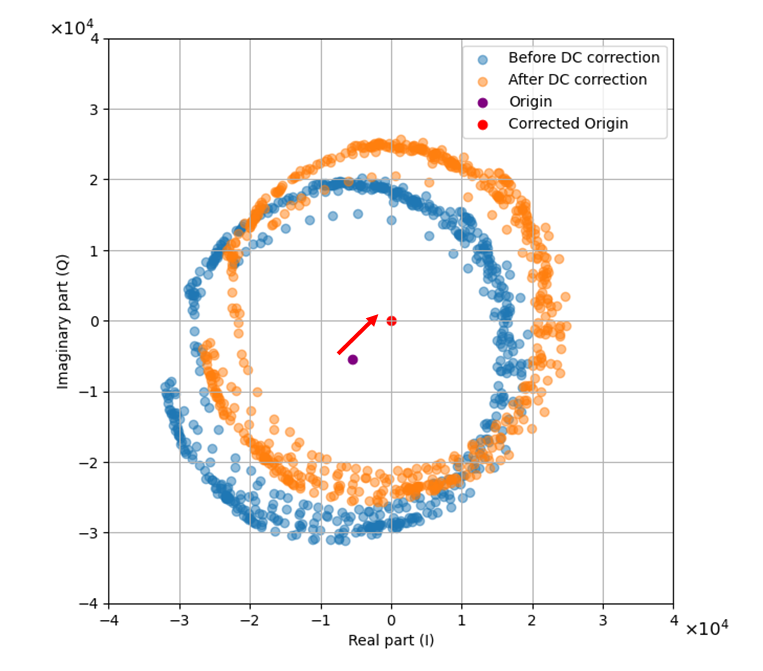}

\caption{I and Q samples constellation before and after DC offset correction.}
\label{DC-offset}
\end{figure}

Figure \ref{DC-offset} illustrates the detrimental effect of DC offset on the signal constellation within the real-imaginary (I and Q) plane. It further demonstrates how our optimization approach, coupled with the Gradient Descent (GD) algorithm, identifies and corrects the DC offset, effectively repositioning the constellation to its original state without the DC offset influence.



\subsection{Noise removal and phase enhancement
}

\subsubsection{Phase Extraction}
After selecting the target range bin associated with the human target and performing DC offset correction, we obtain filtered and clean I and Q values from the sampled signal. As discussed earlier, the displacement of the chest—encompassing both heart rate and respiration rate signals—can be observed in the phase of the received signal.

Applying a simple arctangent operation on the IQ samples yields phase values within the range of $[-\pi, \pi]$. However, given the characteristics of the signal, it's important to recognize that the actual phase values can exceed this range. Chest displacement due to heartbeat and respiration (cardiopulmonary displacements) can reach approximately 12 mm, which significantly surpasses the wavelength of mm-wave FMCW radar. Considering the relationship between phase changes and cardiopulmonary displacements: 

\begin{equation}
    \phi(t)= \frac{4\pi x(t)}{\lambda}
\end{equation}

where $x(t)$ represents the cardiopulmonary chest displacements and $\lambda$ is the wavelength, it is evident that phase values can extend beyond the $[-\pi, \pi]$ range.  Consequently, relying solely on the arctangent for phase calculation leads to phase wrapping issues.

To address these issues, two primary approaches are commonly employed, which we will discuss in the following sections. These methods aim to correctly unwrap the phase, ensuring accurate measurement of chest displacement and, by extension, precise detection of heart and respiration rates.
\paragraph{Method 1. Arc tangent and phase unwrapping}
After calculating the phase signal using the arctangent, the next critical step is phase unwrapping. The initial phase calculation for each sample is given by:

\begin{equation}
    \phi[n]= \arctan  \left( \frac{Q[n]}{I[n]} \right)
\end{equation}

However, this calculation can lead to a discontinuous phase signal due to the periodic nature of the arctan function, resulting in abrupt phase shifts of $2\pi$ radians (a phenomenon known as phase wrapping). Phase unwrapping aims to correct these discontinuities to recover a continuous phase signal. It involves comparing consecutive phase values and adjusting them by adding or subtracting multiples of $2\pi$ whenever a jump 
is detected. This process ensures the phase signal accurately reflects the true, continuous phase changes over time.
\paragraph{Method 2. EDACM}

Despite our efforts to significantly reduce various types of noise through different signal processing techniques, traditional phase unwrapping methods can be inadequate for radar applications. Their vulnerability to noise and rapid phase changes poses a challenge, as they may fail to distinguish between genuine phase shifts and those induced by noise, potentially compromising the accuracy of vital sign detection.

To address these limitations, the discrete antenna-cross correlation method (DACM) is introduced \cite{EDACM}. This approach modifies the traditional method by substituting the arctangent operation with a derivative operation. Below, we present the DACM algorithm and its refined version, the Extended DACM (EDACM), which is a discrete representation of DACM:

\begin{equation}
\frac{d}{dt} [\phi(t)] = \frac{d}{dt} \left[ \arctan \left( \frac{Q(t)}{I(t)} \right) \right] = 
\frac{I(t)Q'(t) - I'(t)Q(t)}{I(t)^2 + Q(t)^2},
\end{equation}

\begin{equation}
\phi[n] = \sum_{k=2}^{n} \frac{I[k]\{Q[k] - Q[k-1]\} - \{I[k] - I[k-1]\}Q[k]}{I^2[k]+ Q^2[k]}.    
\end{equation}

In these equations, $I'(t)$ and $Q'(t)$ represent the derivatives of I and Q, while $I[k]$ and $Q[k]$ denote their sampled values. It's important to note that $\phi[1]$ is undefined in this formulation. In practice, we address this by initializing it with an estimate derived from the arctangent operation and subsequent phase unwrapping.

\begin{figure}[t]
    
\includegraphics[width=9 cm]{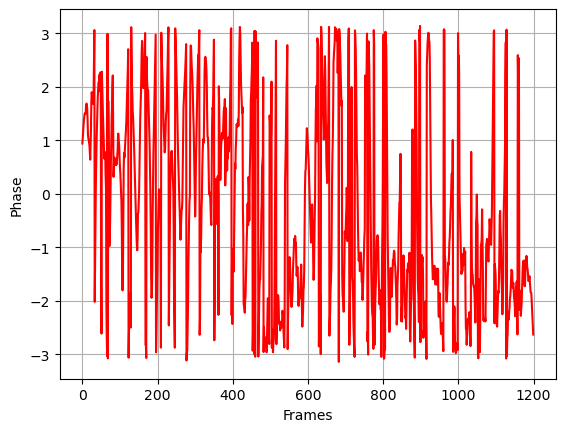}
    
\caption{Phase Values Calculated Using Arc tangent.}
\label{arctan_phase}
\end{figure}

\begin{figure}[h]
    
\includegraphics[width=9 cm]{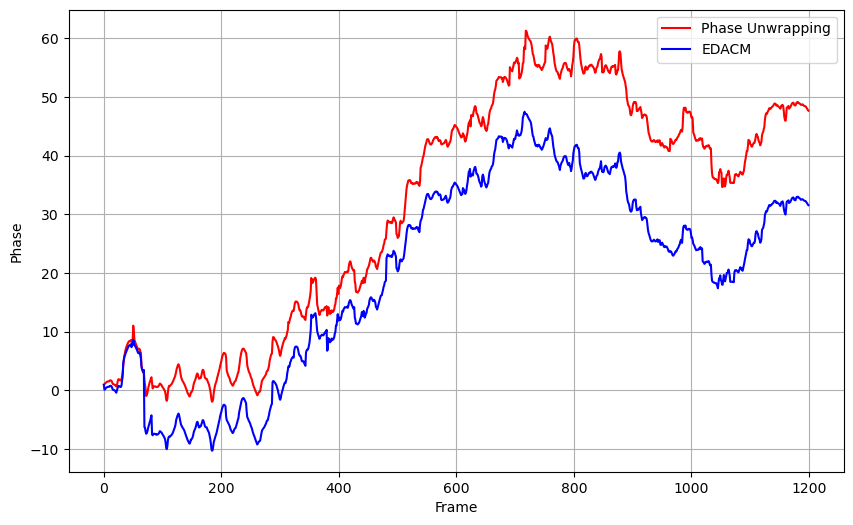}
    
\caption{ Phase Values Following Phase Unwrapping and Application of the EDACM Algorithm. }
\label{comparison}
\end{figure}

In Fig. \ref{arctan_phase}, the calculated phase following the arctangent operation is depicted. It is evident that the estimated values are wrapped and confined within the $[-\pi, \pi]$ range. In Fig. \ref{comparison}, both phase calculation methods are illustrated for comparison. The lack of significant differences indicates that issues related to noise in the phase unwrapping process do not substantially affect the data.

In the results section, we perform a grid search to evaluate how the integration of noise reduction techniques with these two methods influences our signal processing and accuracy estimation. This allows us to identify the most effective approach.

\subsubsection{Phase Difference}


Upon acquiring the phase signal (or equivalently, its samples, $\phi[n]$), our objective is to denoise and amplify this signal to more accurately estimate chest displacement, which encapsulates the effects of both heart and respiratory rates. To achieve this enhancement, we employ a technique that focuses on the differences between consecutive phase samples, rather than the samples themselves. This approach is articulated as follows:

\begin{equation}
    \phi'[n]=\phi[n]-\phi[n-1]
\end{equation}
where $\phi'[n]$ represents the modified phase signal (phase difference), which consequently contains one fewer sample. The motivation behind this strategy includes noise reduction, highlighting vital signs, and the enhancement of heart rate signals, which we elaborate on in the following.

First, this method effectively mitigates low-frequency noise and drift, which are prevalent in phase measurements. In our quest to isolate the chest displacement signal and, by extension, the movements related to heart and respiratory rates, we focus predominantly on temporal changes in the signal. The computation of differences between successive phase samples serves to underscore these temporal variations, rather than the absolute phase values. This emphasis is particularly beneficial, as noise typically has a more pronounced effect on absolute phase measurements, whereas temporal changes are less susceptible.

Secondly, the physiological movements indicative of heartbeats and respiration are subtle and low in amplitude. By focusing on phase changes, we target those signal components that most likely reflect these movements, thereby facilitating their identification against the background noise.


Lastly, although the heart rate exhibits a higher frequency than the respiratory rate, the chest displacement attributed to the cardiovascular system is generally lesser than that caused by respiration and might even be obscured by it. Since phase difference emphasizes changes over time, it can more effectively highlight the faster, more subtle fluctuations associated with heartbeats compared to the slower, larger movements of respiration.
Additionally, the larger, dominant movements associated with respiration can cause significant baseline shifts in the radar signal. The phase difference technique normalizes these shifts by concentrating on changes between consecutive measurements, which helps to diminish the overwhelming influence of large but slow-moving signals like respiration, thus enhancing the detection of heart rate and cardiovascular-related movements within our signal.

\begin{figure}[t]
\includegraphics[width=8 cm]{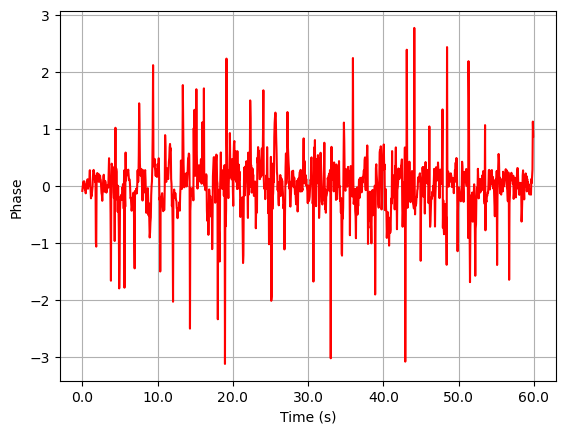}
\caption{Phase signal after applying the phase difference method. }
\label{phase-dif}
\end{figure}

Figure \ref{phase-dif} illustrates the phase signal post-phase difference application. A comparison between Fig \ref{phase-dif} and Fig \ref{comparison} (which presents the phase signal before the phase difference operation) underscores the significance of this technique.

\subsubsection{Impulsive Noise Removal}

While the phase difference method primarily aims to accentuate physiological signals by attenuating slow-changing noise or drift (such as gradual target movements or other slowly varying environmental factors), the phase signal often retains a significant amount of noise, including impulsive noises.
Impulsive noises, characterized by sudden, short spikes in the signal, can mask the underlying physiological signals. These disturbances, noted for their high amplitude and brief duration, may arise from various sources, including electronic interference, sudden movements, or other transient environmental phenomena.


To address impulsive noises within the phase signal, various filtering techniques were explored, including the Median filter, Moving Average filter (MA), Weighted Moving Average filter (WMA), and Exponentially Weighted Moving Average filter (EWMA). Among these, the Median filter and EWMA were found to be particularly effective.
Median Filter generally replaces each sample with the median of neighboring samples within a specified window. As the Median value is less sensitive to extreme values than the mean it is an efficient solution against impulsive noise. Moreover, it helps preserve the edges of the signal while reducing and suppressing noise spikes.
EWMA, an infinite impulse response filter, assigns diminishing weights to older samples. This method is adept at smoothing the signal and diminishing the influence of outliers. Contrary to the Median filter, EWMA offers a nuanced balance between smoothing the signal and maintaining its sensitivity to changes. This balance is vital for preserving the dynamic properties of the signal, essential for the accurate detection of minute variations associated with respiration and heart rates.

\begin{figure}[t]
\includegraphics[width=9cm]{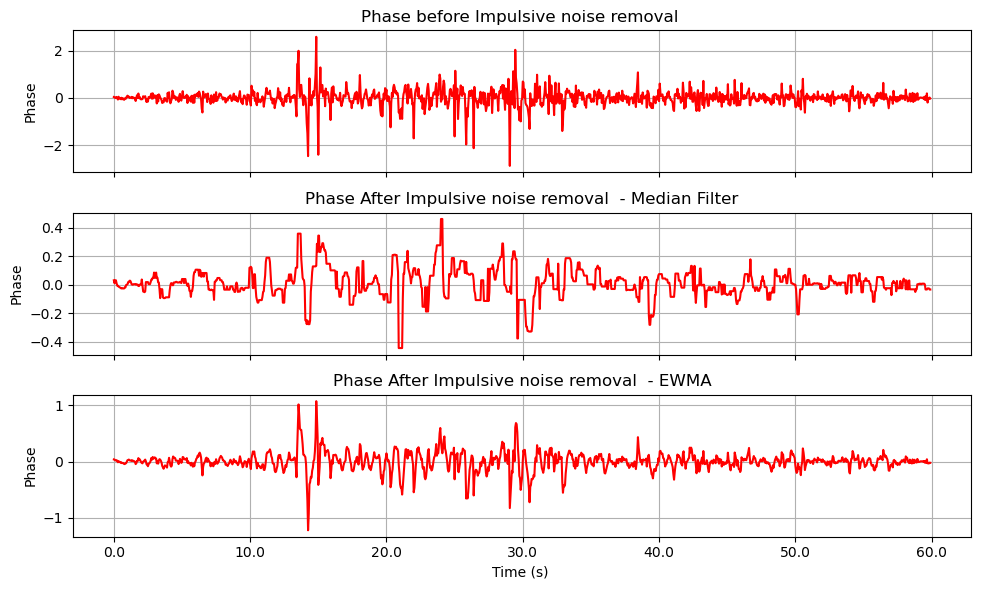}
\caption{The phase signal before and after impulsive noise removal using Median filter and EWMA}
\label{impRMV}
\end{figure}

Figure \ref{impRMV} displays the original phase signal alongside its post-filtered versions using both the Median and EWMA filters. The comparative analysis clearly demonstrates a more refined signal post-filtering, with a notable reduction in the intensity and prevalence of spikes.

\subsection{RX Channels Data Confusion
}
In the development of our FMCW radar system, we have expanded the typical configuration from a single-transmitter, single-receiver setup to a more complex two-transmitter, four-receiver array. This enhancement allows us to capture a broader range of signal reflections and increase the robustness of our measurements. Specifically, before performing bandpass filtering to isolate heart rate and respiration rate frequencies, we introduce a crucial preprocessing step termed ``RX Channels Data Fusion." In this process, we compute the average of the phase signals—or equivalently, the chest displacement signals—across all four receiver channels. This averaging method is not merely a procedural necessity; it plays a critical role in enhancing the signal-to-noise ratio (SNR) of our system. By aggregating the data from multiple channels, we effectively reduce random noise variations and other anomalies that can obscure the physiological signals of interest. Consequently, this fusion not only strengthens the detectability of the heart and respiration rate signals but also improves the overall reliability and accuracy of our health monitoring radar system.
\subsection{Band Pass Filtering}
Once we obtain the cleanly filtered phase signal, equivalent to the chest displacement signal, we can proceed with the extraction of heart rate and respiration rate.
Typically, heart rate signals are found in the 0.7 to 2 Hz range, correlating to 42 to 120 beats per minute, a common range for resting adults. Respiration rates, on the other hand, typically fall within the 0.1 to 0.5 Hz range, equivalent to 6 to 30 breaths per minute. Given these closely spaced frequency bands, filtering becomes an essential step to mitigate interference and isolate the desired physiological signals accurately.

To do so, we employ Butterworth and elliptic filters, two types of Infinite Impulse Response (IIR) filters. 
The Butterworth filter is ideal for keeping HR and RR signal amplitudes undistorted due to its flat passband, ensuring accurate vital sign monitoring. Its predictable, ripple-free response aids in straightforward signal analysis. However, its slower transition may challenge the separation of closely spaced HR and RR frequencies.
The elliptic filter stands out for its quick transition between passband and stopband, crucial for distinguishing between HR and RR signals. Its precision in selecting frequency components comes with the downside of passband and stopband ripples, possibly affecting amplitude accuracy.

\begin{figure}[h]
    \centering
    \begin{subfigure}[b]{0.45\textwidth}
        \includegraphics[width=\textwidth]{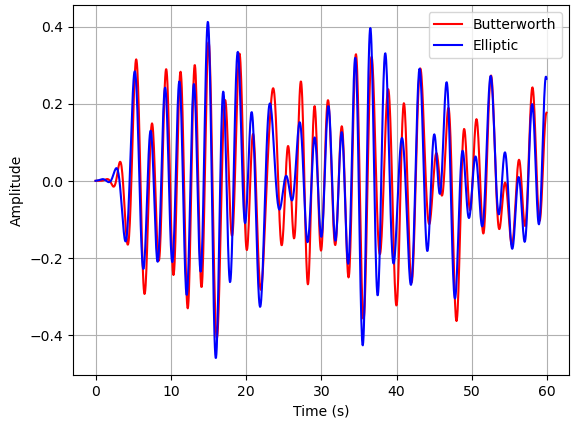}
        \caption{Respiration Rate (RR) Extraction}
        \label{fig:sub12}
    \end{subfigure}
    \hfill 
    \begin{subfigure}[b]{0.45\textwidth}
        \includegraphics[width=\textwidth]{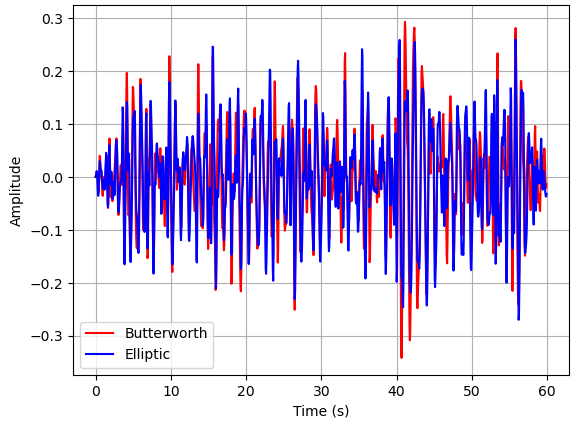}
        \caption{Heart Rate (HR) Extraction}
        \label{fig:sub22}
    \end{subfigure}
    \caption{Comparative Analysis of Respiration Rate (RR) and Heart Rate (HR) Extraction Using Butterworth and Elliptic Filters.}
    \label{BPF}
\end{figure}

Considering all these, Utilizing both Butterworth and elliptic filters allows for exploiting their strengths—signal integrity with Butterworth and selective separation with the elliptic filter. Testing both provides a thorough performance evaluation, guiding the optimal filtering choice for precise HR and RR extraction in remote vital sign monitoring.
That is why, in the result section we explore both filters. Finally, Figure \ref{BPF} displays the chest displacement signal after undergoing bandpass filtering with both Elliptic and Butterworth filters.

\section{Heart Rate and Respiration rate Extraction Methods}

\begin{figure}[h]
\includegraphics[width=8.7cm]{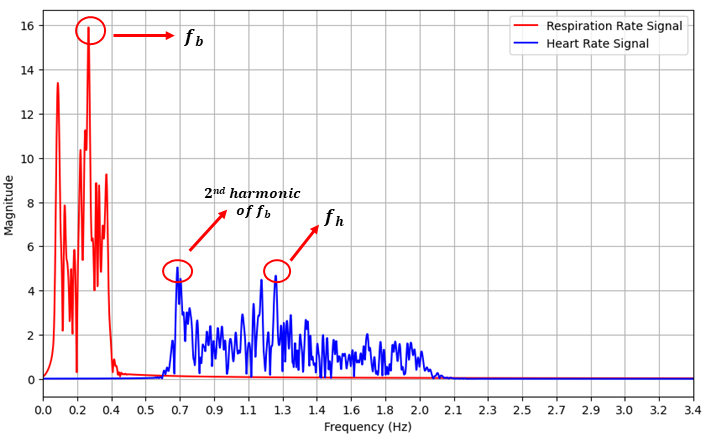}
\caption{Frequency domain representation of filtered heart rate and respiration rate signals, highlighting fundamental and second harmonic peaks.}
\label{harmonics}
\end{figure}

In heart rate respiration extraction, we face complex signal processing challenges, especially due to the interference of respiratory harmonics in the heart rate signal. Vital signs are typically extracted from radar signals based on their respective frequency components; however, the respiration rate is not strictly periodic. The amplitude and frequency of respiratory signals vary with physiological and environmental factors, leading to a complex spectral presence with significant higher harmonics due to the large chest displacement during respiration compared to that of heartbeats.
These harmonics often overlap with the frequency band used for heart rate analysis, potentially obscuring the true heart rate signal, which complicates its accurate detection. Additionally, the non-cyclostationary nature of the respiration signal—where mean, variance, and periodicity vary—further complicates the situation by generating multiple harmonic frequencies.
Our findings indicate that typically, only the second harmonics have a magnitude comparable to that of the heart rate. Fig. \ref{harmonics} displays the frequency domain representation of filtered signals. Here, the dominant peak in the respiration rate signal is the respiration frequency, while in the heart rate signal, the second harmonics of the respiration rate and the heart rate display approximately equal magnitudes, which can lead to incorrect heart rate frequency estimation.

This complexity necessitates the development of more sophisticated heart rate and respiration extraction methods. Consequently, we have implemented and developed various extraction techniques to address these challenges.

\subsubsection{Improved FFT}
As an initial method to identify the dominant frequencies in filtered heart rate and respiration rate signals, we employ the FFT. Traditional methods like FFT often struggle in noisy environments or when the signal contains closely spaced frequency components, such as the harmonics of respiration rates found in heart rate signals. This challenge necessitates the exploration of more sophisticated spectral estimation methods, which we will discuss subsequently.
The resolution of FFT is defined by the equation:

\begin{equation}
    f_{res}=\frac{f_{s}}{N},
\end{equation}
where $f_{s}$ represents the sampling frequency and $N$ the length of the signal, respectively. In our case, this results in a resolution of 0.03 Hz for a sampling frequency of 20 Hz and a signal length of 600. Since the sampling frequency is fixed and tied to the frame duration, we can enhance the frequency resolution by increasing the length of the signal. One way to achieve this is through zero padding, which effectively extends the signal length and improves the frequency resolution in FFT. We refer to this enhanced method as ``Improved FFT" throughout the remainder of the paper.

Moreover, as heart rate estimation is susceptible to errors due to the presence of the second harmonic of the respiration rate, we also implement a Notch filter to remove this second harmonic from the heart rate signal spectrum in the heart rate extraction step.
This is achieved by using the respiration rate frequency estimated with the Improved FFT method described earlier.

\subsubsection{Coarse To Fine Estimation (CTF)}

While other algorithms are employed for extracting both heart and respiration rates, the coarse-to-fine estimation approach stands out as a sophisticated method specifically designed to accurately determine heart rate from FMCW radar data. Initially, a coarse estimation of the heart rate frequency is made using time-domain data; this is then refined using frequency-domain analysis to achieve precise results, referred to as fine estimation. This method was first proposed by K. Liu et al. (2020) \cite{myref-1}.

\begin{figure}[t]
\includegraphics[width=8.5 cm]{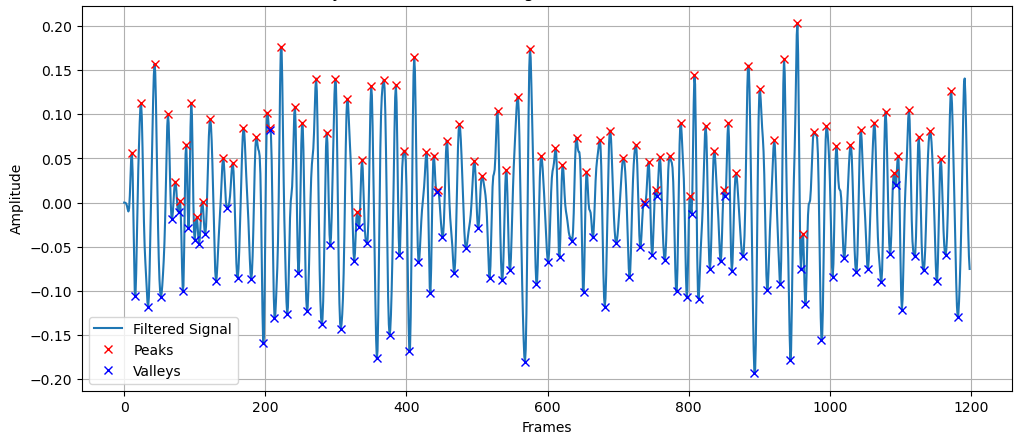}
\caption{Detected peaks and valleys in the filtered heart rate signal.}
\label{pvpair}
\end{figure}

During the coarse estimation phase, the method involves detecting peaks and valleys in the chest displacement signal derived from the radar data. These peaks and valleys are critical as they represent the fundamental heart rate signal amidst other bodily motion artifacts. The distances between these peak-valley pairs are calculated and analyzed to distinguish between potential heart rate signals and noise. Figure \ref{pvpair} depicts the peaks and valleys detected in the filtered heart rate signal. Subsequently, two methods are employed for coarse estimation.

\begin{figure}[t]
\includegraphics[width=7 cm]{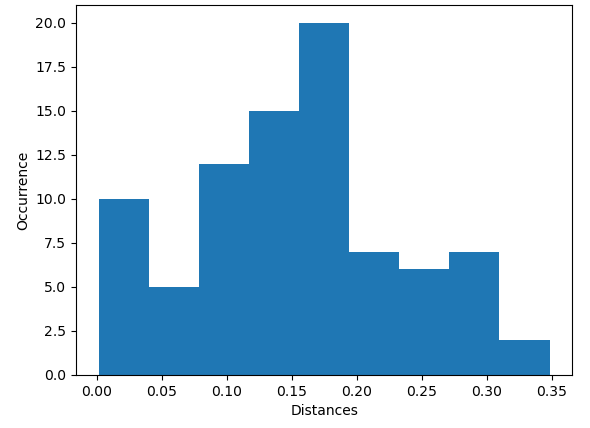}
\caption{Histogram of peak-valley distances in the heart rate signal.}
\label{his}
\end{figure}

\begin{figure}[t]
\includegraphics[width=8.5 cm]{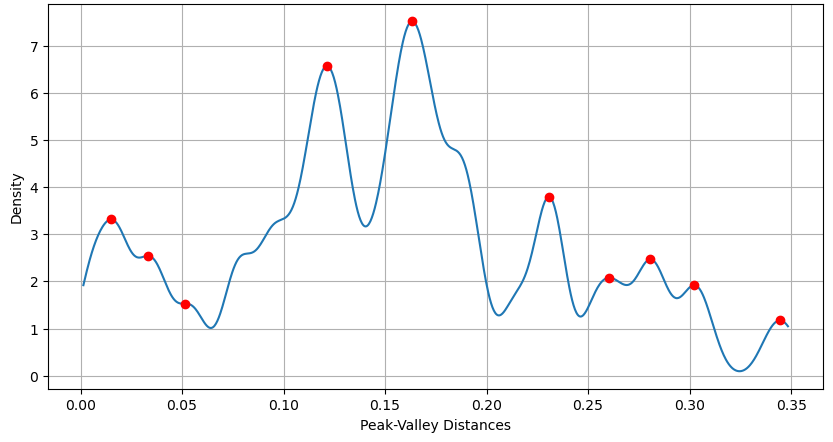}
\caption{Kernel Density Estimation of peak-valley distances for heart rate analysis. }
\label{kde}

\end{figure}

\paragraph{Coarse Estimation- Histogram-Based Approach }
This method sets an adaptive threshold based on the distribution of peak-valley distances, initially determined by the 25th percentile of these measurements. Distances below a certain multiple of this threshold are considered noise, as shorter peak-valley distances typically indicate noise interference. This approach aids in identifying the valid peak-valley pairs that are significant for heart rate detection, corresponding to the clear peaks in the histogram and providing a coarse estimation of heart rate frequency. The histogram, generated from peak-valley pairs, is shown in Fig. \ref{his}. In subsequent sections, the method will be referred to as CTF-His.

\paragraph{Coarse Estimation- Kernel Density Estimation (KDE) Approach}
Kernel Density Estimation (KDE) offers a more refined way to analyze the distribution of peak-valley distances. By estimating the probability density function of these distances, KDE helps identify the most prominent distances likely associated with the actual heart rate while effectively filtering out noise. KDE provides enhanced accuracy over histogram-based methods by smoothing out noise, allowing bandwidth customization, and capturing complex signals more accurately. This makes KDE particularly effective in a coarse-to-fine estimation framework for reliably detecting heart rate amidst bodily noise using FMCW radar. The KDE result is displayed in Fig. \ref{kde}, showing the distribution of peak-valley distances. In subsequent sections, the method will be referred to as CTF-KDE.

The coarse heart rate frequency is then calculated by counting the number of valid peak-valley pairs over the total signal time, providing an initial estimate of heart rate in beats per minute (bpm).

\paragraph{Fine Estimation}

Following the coarse estimation, an FFT is applied to the same heart rate signal to transform it into the frequency domain. This transformation reveals the spectrum of frequencies present in the signal, allowing for precise identification of the heart rate frequency.
The peaks in the FFT are ranked by their amplitudes, and the frequency closest to the coarse estimation is selected as the fine estimation of the heart rate. This method enhances the accuracy by verifying the time-domain analysis with frequency-domain data.

\subsubsection{MUSIC}

The Multiple Signal Classification (MUSIC) algorithm is a sophisticated technique in signal processing that leverages the eigenstructure of the covariance matrix of sensor array data to estimate the directions of arrival (DoA) of signals. This approach, pioneered by R. O. Schmidt in the 1980s, has since become a fundamental tool in the fields of radar, sonar, and telecommunications, among others. Its main application lies in the ability to resolve closely spaced signals in scenarios where traditional methods struggle due to limited resolution. Although it has been predominantly utilized for DoA estimation, in this work, we introduce a significant modification by applying the MUSIC algorithm to extract respiration rate and heart rate from chest displacement signals captured by radar. This adaptation allows the MUSIC algorithm to be applied in analyzing the frequency components of physiological signals, a novel use case that extends its applicability beyond its conventional domain. This process involves several key steps: Covariance Matrix Estimation, Eigenvalue Decomposition, and MUSIC Spectrum formation.

\paragraph{Covariance Matrix Estimation}
Consider the chest displacement signal (phase signal) after undergoing band-pass filtering for heart rate and respiration rate detection. This filtered signal can be represented as x(n), where n indexes the time samples.  The covariance matrix, denoted as $\mathbf{R}$, is calculated using the formula:

\begin{equation}
\mathbf{R} = \frac{1}{N} \sum_{n=1}^{N} \mathbf{x}(n) \mathbf{x}^{H}(n),
\end{equation}
where \( \mathbf{x}^{H}(n) \) is the Hermitian transpose (conjugate transpose)
of \( \mathbf{x}(n) \), and \( N \) is the number of samples which depends on the observation time. 

The covariance matrix reflects how these signals are correlated and captures how the signal varies over time.
In our adaptation, we use lagged versions of the signal to build the covariance matrix. This method helps us understand how the signal changes over time and find patterns within it more efficiently by capturing the temporal structure and correlation within the signal. By doing this, we lay a good groundwork for the next steps, which are Eigenvalue Decomposition and creating the MUSIC Spectrum. This careful way of making the covariance matrix not only helps us figure out what frequencies are in the signal but also improves how well the MUSIC algorithm can identify and pull out vital signs like heart rate and respiration rate from the signals we get from radar.

\paragraph{Eigenvalue Decomposition}
In this step, we decompose the covariance matrix $\mathbf{R}$ into signal and noise sub-spaces. 
The eigenvectors corresponding to the largest eigenvalues form the signal subspace, encapsulating the main frequencies of interest in our signal—such as those from heart or respiration rates. The remaining eigenvectors, associated with smaller eigenvalues, define the noise subspace, representing elements of the signal considered as noise or irrelevant information. 
The distinction between these subspaces is crucial and is typically determined based on the expected number of signal sources, indicating the dominant frequencies present in the signal. This decomposition process is mathematically represented as:

\begin{equation}
\mathbf{R} = \mathbf{E}_s \mathbf{\Lambda}_s \mathbf{E}_s^H + \mathbf{E}_n \mathbf{\Lambda}_n \mathbf{E}_n^H.
\end{equation}

In this equation,  $\mathbf{E}_s$ and $\mathbf{E}_n$
denote the matrices of eigenvectors corresponding to the signal and noise subspaces, respectively. The matrices 
$\mathbf{\Lambda}_s$ and $\mathbf{\Lambda}_n$
are diagonal matrices filled with the eigenvalues of the covariance matrix $\mathbf{R}$, divided into components related to the signal subspace and the noise subspace, respectively.

For the estimation of respiration rate, the signal subspace is expected to primarily capture the respiration rate signal. This is because, in such analyses, the respiration rate signal often presents the most significant periodic component detectable in chest displacement signals.
Conversely, for heart rate estimation, the signal subspace should encompass not only the heart rate signal but also the second harmonic of the respiration rate. This inclusion ensures that both primary cardiac rhythms and significant respiratory effects are accounted for, enhancing the precision of heart rate measurements.
 
The noise subspace, composed of eigenvectors associated with the smallest eigenvalues, is orthogonal to the signal subspace. It encompasses elements of the signal considered as noise, including unrelated harmonics of respiration rate and other external noise sources. This orthogonality ensures that the signal subspace is clearly separated from noise, allowing for a more accurate extraction of vital signs from the analyzed signals.

\paragraph{MUSIC Spectrum}
Finally, we adapt the MUSIC spectrum, originally designed for Direction of Arrival (DoA) estimation varying by $\theta$ (degree), to extract dominant frequencies of the signal for heart rate and respiration rate estimation. Consequently, the MUSIC spectrum for these vital signs can be expressed as:

\begin{equation}
P_{\text{HR}}(f) = \frac{1}{\mathbf{v}_{HR}^H(f) \mathbf{E}_{HR,n} \mathbf{E}_{HR,n}^H\mathbf{v}_{HR}(f)},
\end{equation}

\begin{equation}
P_{\text{RR}}(f) = \frac{1}{\mathbf{v}_{RR}^H(f) \mathbf{E}_{RR,n} \mathbf{E}_{RR,n}^H\mathbf{v}_{RR}(f)}.
\end{equation}

Here, $v(f)$ represents the steering vector, which, unlike in DoA applications where it varies with the angle $\theta$, is defined based on different frequency contents of the signal for our use case. This adaptation leverages the orthogonality principle between the signal and noise subspaces, enabling the detection of frequencies contained within the steering vectors.
In contrast to the angle-shifted signal series used in DoA estimation, we define the steering vector for our purposes as follows:

\begin{equation}
\mathbf{v}(f) = \left[ 1, e^{-j2\pi \frac{f}{f_s}}, e^{-j2\pi 2\frac{f}{f_s}}, \ldots, e^{-j2\pi (N-1)\frac{f}{f_s}} \right]^T
\end{equation}

In this formulation, \(f\) denotes the frequency at which the MUSIC spectrum is being evaluated, \(f_s\) is the sampling frequency of the signal, and \(N\) is the number of samples or the effective length of the steering vector. It is important to note that using a lagged signal may influence these parameters.
 When the MUSIC algorithm calculates the spectrum, it exploits this orthogonality. 

The MUSIC algorithm exploits the orthogonality while calculating the spectrum.
When the components of the steering vector correspond to a valid frequency within the signal, they will be orthogonal to the noise subspace. As a result, peaks in the MUSIC spectrum occur at frequencies where the steering vector is orthogonal to the noise eigenvectors, revealing the signal's frequency content (either heart rate or respiration rate). This approach underscores the MUSIC algorithm's flexibility and effectiveness in identifying vital sign frequencies from radar-derived phase signals.

\begin{figure}[t]
    \centering
    \begin{subfigure}[b]{0.5\textwidth}
        \includegraphics[width=\textwidth]{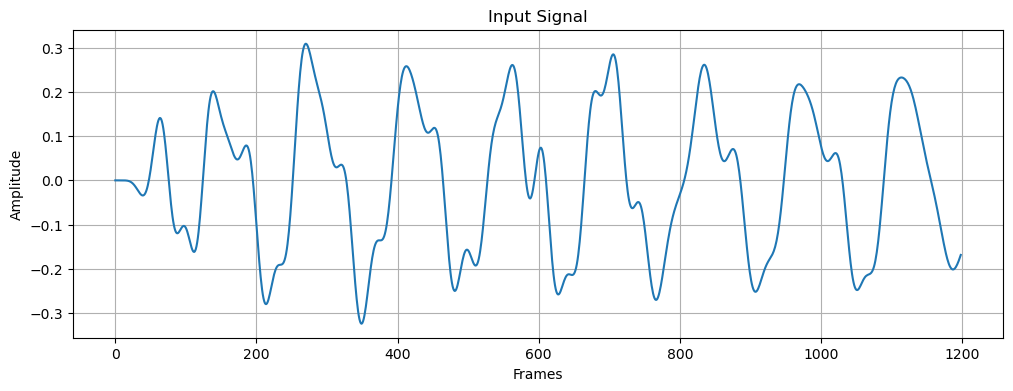}
    \end{subfigure}
    \hfill 
    \begin{subfigure}[b]{0.5\textwidth}
        \includegraphics[width=\textwidth]{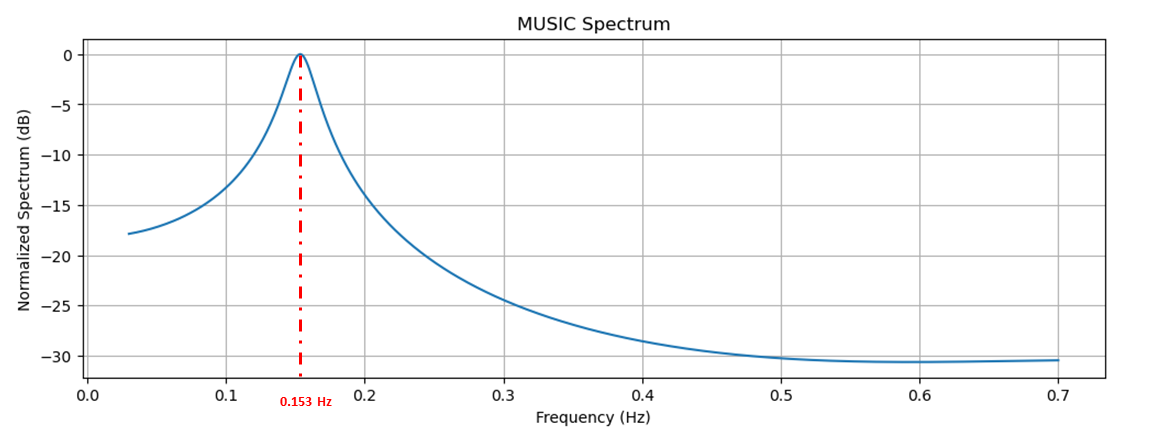}
    \end{subfigure}
    \caption{Filtered respiration rate Signal and its MUSIC Spectrum for Frequency Estimation.}
    \label{MUSIC-RR}
\end{figure}

The algorithmic overview of the steps is outlined in Algorithm \ref{alg1}. It should be noted that the algorithm is run separately for heart rate and respiration rate detection because they comprise distinct frequency components. In Fig. \ref{MUSIC-RR}, the input signal (filtered respiration signal) and the MUSIC spectrum are displayed. For this measurement, the reference respiration rate was 9 RPM, and the MUSIC spectrum indicates a peak at 0.153 Hz, corresponding to an estimated rate of 9.211 RPM.

\begin{figure}[h]
    \centering
    \begin{subfigure}[b]{0.5\textwidth}
        \includegraphics[width=\textwidth]{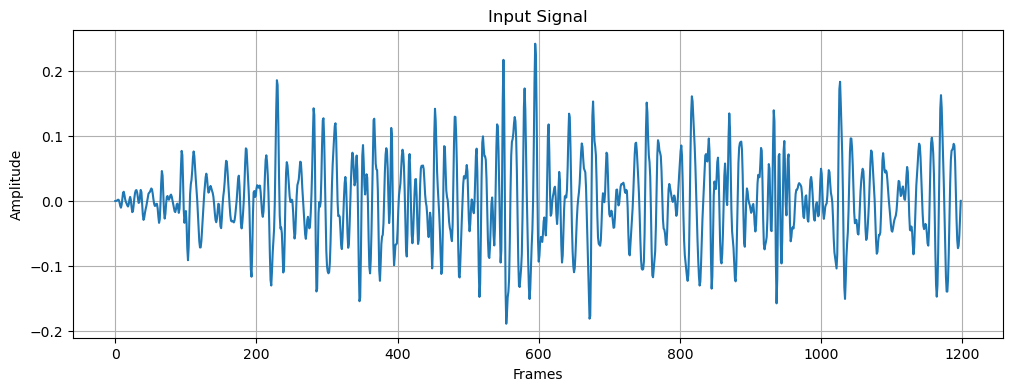}
    \end{subfigure}
    \hfill 
    \begin{subfigure}[b]{0.5\textwidth}
        \includegraphics[width=\textwidth]{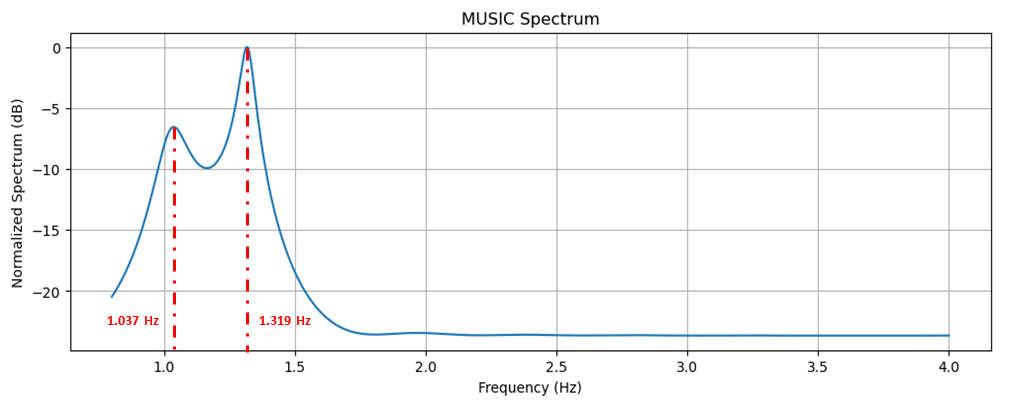}
    \end{subfigure}
    \caption{Filtered Heart Rate Signal and its MUSIC Spectrum for Frequency Estimation.}
    \label{MUSIC-HR}
\end{figure}

In Fig. \ref{MUSIC-HR}, the input signal (filtered heart signal) and the MUSIC spectrum are presented. The spectrum for heart rate differs from that of the respiration rate, often showing two peaks. The dominant peak corresponds to the heart rate, estimated at 1.319 Hz, which translates to 79.166 BPM. This is in close agreement with the reference average heart rate of 78.2 BPM for the observation period. The secondary peak, occurring at 1.037 Hz (62.23 BPM), is likely one of the harmonics of the respiration rate, possibly the 6th or 7th harmonic.

\begin{algorithm}

\caption{MUSIC Algorithm for Vital Sign Detection}
\begin{algorithmic}[1]

\State \textbf{Input:} Chest displacement (phase) signal after bandpass filtering (heart rate or respiration rate)
\State \textbf{Output:} Estimated frequencies of heart rate and respiration rate

\Procedure{Vital Signs Detection}{}

\State Do the following steps for each observation window
\State Determine the lag to be applied to the signal.
\State Calculate the covariance matrix $\mathbf{R}$.
\State Perform eigenvalue decomposition on $\mathbf{R}$.
\State Sort the eigenvalues in descending order.
\State Split eigenvectors into signal ($\mathbf{E}_s$) and noise ($\mathbf{E}_n$)
\Statex \hspace{11pt} subspaces, based on the number of expected 
\Statex \hspace{11pt} signal sources.

\State Specify the frequency range of interest for heart 
\Statex \hspace{11pt} rate and respiration rate.
\State Construct the steering vector $\mathbf{v}(f)$ for 
\Statex \hspace{11pt} the specified frequency range.
\State Define the resolution and compute the MUSIC 
\Statex \hspace{11pt} spectrum $P_{\text{HR}}(f)$ and $P_{\text{RR}}(f)$.
\State Identify peaks in the MUSIC spectrum.

\State \Return \parbox[t]{\dimexpr\linewidth-\algorithmicindent}{Estimated frequencies.}
    
\EndProcedure

\end{algorithmic}\label{alg1}
\end{algorithm}

\subsubsection{Prony}

The Prony method, developed by Gaspard Riche de Prony in the 18th century, is an analytical tool designed to decompose a signal into its sinusoidal components, modeling it as a sum of damped sinusoids. This method proposes that a signal $x[n]$  can be represented as:

\begin{equation}\label{p-1}
     x[n] = \sum_{k=1}^{M} A_k e^{(j\omega_k + \alpha_k)n} + noise.
\end{equation}
where \( M \) is the number of sinusoidal components, \( A_k \) is the amplitude, \( \omega_k \) is the angular frequency, \( \alpha_k \) is the damping factor, \( n \) is the time index, and noise represents background noise or interference.
In this formulation \( A_k \) may be complex, allowing the expression of $x[n]$ to also be written as:

\begin{equation}
  x[n] \approx \sum_{k=1}^{M} A_k e^{\alpha_k n} \left( \cos(\omega_k n + \phi_k) + j \sin(\omega_k n + \phi_k) \right),
\end{equation}
Where $ A_k $ and $\phi_k$ represent the amplitude and phase of each sinusoidal component, respectively. This representation is particularly advantageous for analyzing complex signals characterized by multiple overlapping components. It enables the estimation of the amplitude, frequency, and phase of each component, providing a comprehensive understanding of the signal's composition.

Once the parameters of the Prony representation are determined, we can discern the frequency components constituting the filtered phase signal (equivalent to chest displacement), and the magnitude of each component indicates its dominance. To achieve this understanding, the method involves several critical steps, which will be detailed subsequently. 
It's important to note that the signal 
$x[n]$, as mentioned throughout this discussion, specifically refers to the phase signal representative of chest displacement. This signal has undergone band-pass filtering, a crucial preprocessing step aimed at isolating frequency components pertinent to heart rate and respiration rate estimation. For the sake of simplicity and clarity in our exposition, we refer to this filtered phase signal as 
$x[n]$.

\paragraph{Construct the Hankel Matrix}

To initiate the estimation of the signal's component parameters, the first critical step involves constructing a Hankel matrix. This matrix is encapsulating the dynamics of the signal across various time lags, thereby providing a well-organized framework for the subsequent analysis. More importantly, it serves as a foundational step towards computing the linear prediction coefficients, which play a pivotal role in the Prony method. These coefficients are essential for predicting future signal values from past observations and are a key focus of later discussions.

Given a signal $x[n]$ comprising N samples, and with the assumption that our signal model incorporates 
M damped sinusoids, the Hankel matrix is structured to have 
$(N -M+1)\times M$ dimension. The construction of this matrix is designed to facilitate the linear predictive modeling process, as illustrated below:

\begin{equation}
\small
\begin{bmatrix}
x[0] & x[1] & x[2] & \dots & x[M-1] \\
x[1] & x[2] & x[3] & \dots & x[M] \\
x[2] & x[3] & x[4] & \dots & x[M+1] \\
\vdots & \vdots & \vdots & \ddots & \vdots \\
x[N-M-1] & x[N-M] & x[N-M+1] & \dots & x[N-1]
\end{bmatrix}
\end{equation}

\paragraph{Solve the Linear Prediction Equation}

By solving the linear prediction equation, we obtain coefficients ($a_k$) that predict future signal values based on past samples. This step is crucial for identifying the underlying patterns within the noisy radar signals. 
The linear prediction equation is formulated as follows:

\begin{equation}
\hat{x}[n] = -\sum_{k=1}^{M} a_k x[n-k],
\end{equation}
where $\hat{x}[n]$ is the predicted value, $ a_k$ are the prediction coefficients, $ M $ is the number of sinusoidal components, and $ x[n-k] $ are the past values of the signal.
Solving this equation yields the prediction coefficients $ a_k$.
To solve this equation we formulate the prediction error given by:

\begin{equation}
    e[n] = x[n] + \sum_{k=1}^{M} a_k x[n-k].
\end{equation}

This equation can be elegantly expressed in matrix notation, leveraging the Hankel matrix established in the preceding step, as:
\begin{equation}
     H \cdot a = -x'.
\end{equation}

Here $H $ is the Hankel matrix, $ a $ is the vector of prediction coefficients, and $ x'$ is the offset signal vector.
The prediction coefficients are obtained
by solving the matrix equation, typically using a least squares method
to minimize the sum of squares of the errors.
Finding these prediction coefficients is crucial for constructing the companion matrix, which, in subsequent steps, reveals the frequencies and damping factors of the sinusoids.

\paragraph{Find the Roots of the Companion Matrix}

With the prediction coefficients obtained from the previous step, we can now construct the companion matrix, denoted as $C$. 
This matrix is uniquely structured: its last row is filled with the negative values of our linear prediction coefficients
$-a_{1},...,-a_{M}$, while the elements just above the main diagonal—known as the upper diagonal—are all set to 1. This arrangement results in an $M \times M$ square matrix, outlined as follows:

\begin{equation}
     C = \begin{bmatrix}
        0 & 1 & 0 & \ldots & 0 \\
        0 & 0 & 1 & \ldots & 0 \\
        \vdots & \vdots & \ddots & \ddots & \vdots \\
        0 & 0 & 0 & \ldots & 1 \\
        -a_1 & -a_2 & -a_3 & \ldots & -a_M
        \end{bmatrix}
\end{equation}

The next crucial step involves analyzing this matrix to identify its roots, which are key to finding the sinusoidal components hidden within our signal. These components are essential for extracting the frequencies indicative of heart rate and respiration rate.

To do so, we calculate the eigenvalues of the matrix $C$. These eigenvalues are the roots we are looking for, denoted as $ r_k = e^{(\alpha_k + j\omega_k)} $, where $ \alpha_k$  denotes the damping factor, and $\omega_k $ signifies the angular frequency of the $k$-th sinusoidal component. 
It is these angular frequencies that relate directly to the heart rate and respiration rate components in our signal, $x[n]$.
To translate these angular frequencies $\omega$ into the more familiar frequency $f$ measured in Hertz, the following conversion is used:

\begin{equation}
    f = \frac{\omega}{2\pi} = \frac{\omega_k \times f_s}{2\pi}
\end{equation}

\paragraph{Determine Amplitudes and Phases}
While the frequency components of the signal have been identified, it is essential to also determine their amplitudes and phases. This necessity arises from the presence of higher harmonics of the respiration rate within the heart rate signal, leading to multiple frequency components. Therefore, to discern which frequencies are predominant, we must calculate the amplitudes of these components.

The real part of the signal, as represented by the Prony algorithm, can be formulated as:

\begin{equation}
    x[n] \approx \sum_{k=1}^{M} A_k e^{\alpha_k n} \cos(\omega_k n + \phi_k)
\end{equation}

Given the signal values $x[n]$, alongside the identified frequencies and damping factors, we can establish a set of equations. For each sample point $ n $ ranging from $ 0 $ to $ N-1 $, we configure $ N $ equations with $ 2M $ unknown parameters, $A_k $ and $ \phi_k $ for $ k = 1, \ldots, M $. 
Solving these equations yields values for $A_k $ and $ \phi_k $, thus providing a comprehensive Prony representation of the signal.

To find solutions to these equations, the least squares method is employed. This approach aims to minimize the discrepancy between the observed signal values ($x[n]$) and those predicted by our model, focusing on reducing the sum of squared differences:

\begin{equation}
     S = \sum_{n=0}^{N-1} \left( x[n] - \sum_{k=1}^{M} A_k e^{\alpha_k n} \cos(\omega_k n + \phi_k) \right)^2 
\end{equation}

Numerical methods are then applied to solve this set of equations, providing estimates for $A_k $ and $ \phi_k $. 
With these parameters, we can accurately describe our signal as a sum of damped sinusoids. By examining the dominant frequency components in the filtered signals for heart rate and respiration rate, we can accurately report the estimated frequencies for these vital signs.
Finally, an overview of the steps involved in the Prony algorithm is presented in Algorithm \ref{alg-p}.
\begin{algorithm}
\caption{Prony Algorithm for Vital Sign Detection}
\begin{algorithmic}[1]

\State \textbf{Input:} Signal $x[n]$, Number of samples $N$, Model order $M$, Sampling frequency $f_s$
\State \textbf{Output:} Estimated frequencies $f_k$, Damping factors $\alpha_k$, Amplitudes $A_k$, Phases $\phi_k$

\State \textbf{Step 1: Construct the Hankel Matrix}
\State Form Hankel matrix $H$ using signal samples $x[n]$

\State \textbf{Step 2: Solve the Linear Prediction Equation}
\State Solve $H \cdot a = -x'$ for prediction coefficients $a$

\State \textbf{Step 3: Construct the Companion Matrix}
\State Formulate companion matrix $C$ using coefficients $a$

\State \textbf{Step 4: Find the Roots of the Companion Matrix}
\State Calculate the eigenvalues of $C$, denoted as roots $r_k$

\State \textbf{Step 5: Determine Angular Frequencies and Damping Factors}
\State Calculate $\omega_k$ and $\alpha_k$ from the roots $r_k$

\State \textbf{Step 6: Convert Angular Frequencies to Frequencies}
\State $f_k = \frac{\omega_k \times f_s}{2\pi}$

\State \textbf{Step 7: Determine Amplitudes and Phases}
\State Set up equations based on the real part of the signal representation
\State Use the least squares method to solve for $A_k$ and $\phi_k$

\State \textbf{return} $f_k$, $\alpha_k$, $A_k$, $\phi_k$

\end{algorithmic}\label{alg-p}
\end{algorithm}

\section{Experimental Setup}
\subsection{Setup}

In this study, we utilized the AWR1642 EVM mm-wave FMCW radar from Texas Instruments, which operates within a frequency range of 77 to 81 GHz, providing up to 4 GHz of bandwidth \cite{TI-AWR1642EVM}. The radar features two transmitter antennas and four receiver antennas. To maximize the capabilities of the FMCW radar, we employed a MIMO configuration that utilized all available transmitter and receiver antennas. The radar system was complemented by a DCA1000 for raw data collection \cite{TI-DCA}. Both the AWR1642 EVM and the DCA1000 were mounted in a specially designed holder and positioned at approximately 90 cm above the ground.

Radar configurations and parameters are detailed in Table \ref{table-conf}. We utilized the full 4 GHz bandwidth available, which, according to the range resolution formula  $d_{res}=\frac{C}{2BW}$, provided a range resolution of 0.375 meters for our setup. The radar was set to its highest sampling rate, and each frame consisted of 128 chirps to explore the average chirp method.
Furthermore, reference heart rate values were collected using a Polar H10 belt. As the recordings lasted approximately one minute, subjects were asked to count both their inhalation and exhalation to enhance accuracy. The counts reported by the subjects were verified using a high-resolution camera focused on the chest area of the subjects.

\begin{table}
\centering
\begin{tabular}{|c|c|}
\hline
\textbf{Parameter}	& \textbf{Value}\\
\hline
        Start Frequency		& 77 GHz\\
        End Frequency		& 81 GHz\\
        ADC Start time	    & 6 $\mu$s\\
        ADC Samples		    & 250\\
        Sample Rate		    & 6250 ksps\\
        Ramp End Time		& 50 $\mu$s\\
        Idle Time		    & 7 $\mu$s\\
        Frame Count		    & 1200\\
        Frame Periodicity   & 50 ms \\
        Chirp Count    & 128 \\
        RX Gain		        & 30 dB\\
        TX Count		    & 2\\
        RX Count		    & 4\\   
\hline
\end{tabular}
\caption{Configuration parameters of AWR1642 EVM.} \label{table-conf}
\end{table}

\subsection{DataSet, Scenarios and Special Cases}
The dataset used in this study has already been published and is available online, accompanied by a data descriptor paper that thoroughly details the dataset to facilitate its reuse by other researchers \cite{dataset}. This dataset encompasses data from 10 participants (5 males and 5 females) who all engaged in three different scenarios: Distance, Orientation, and Angle.
Our dataset includes a total of 22,560 seconds of recordings obtained through radar.

In the Distance scenario, participants were seated still in front of the radar at distances of 40 cm, 80 cm, 120 cm, and 160 cm, facing the radar. Four measurements, each lasting one minute, were recorded for each participant at these distances.
In the Orientation scenario, participants sat 80 cm away from the radar, facing it with their front, left side, right side, and back in turn. For each orientation, four one-minute recordings were made for each participant.
In the Angle scenario, participants were positioned 80 cm from the radar at angles of 0, 30, and 45 degrees. To test the radar's functionality and our signal processing methods, participants faced a wall rather than the radar, allowing for angled measurements of the chest area. Four one-minute recordings were made for each condition and participant.

Additionally, four participants (participants 2, 3, 4, and 6) took part in the Elevation Scenario, where they were instructed to quickly climb the stairs of a five-floor building. Immediately after this exercise, they were seated 80 cm from the radar, facing it, and two one-minute recordings were captured. Given the rapid decline in heart and respiration rates post-exercise, this exercise was repeated, and two additional recordings were collected.

\begin{figure}[t]
\hspace{-0.4cm}
\includegraphics[width=9 cm]{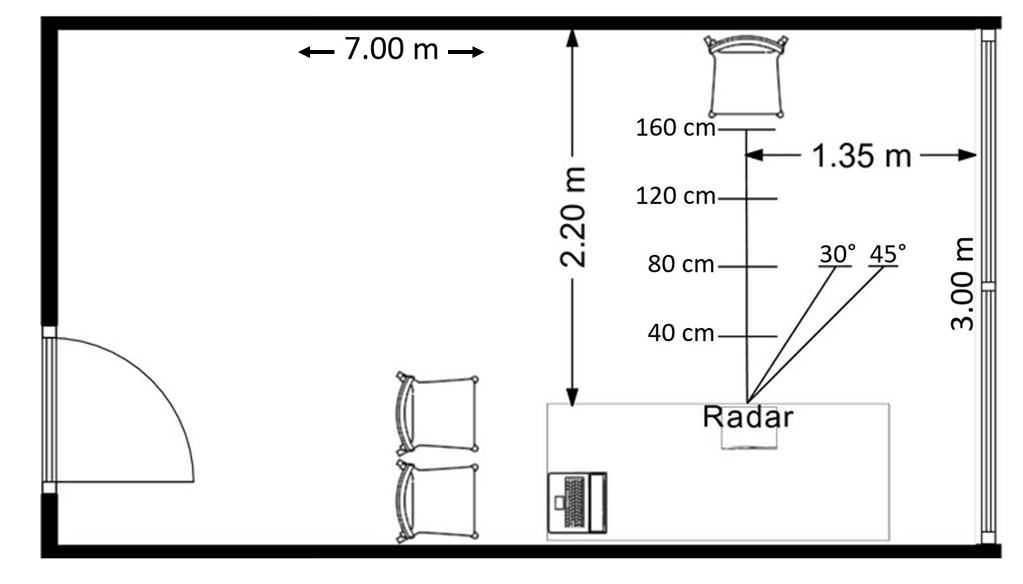}              
\caption{Layout of the experimental setup detailing the radar's position and the marked distances for seating placements.}
\label{map}
\end{figure}

\begin{figure}[t]

\includegraphics[width=8.5 cm]{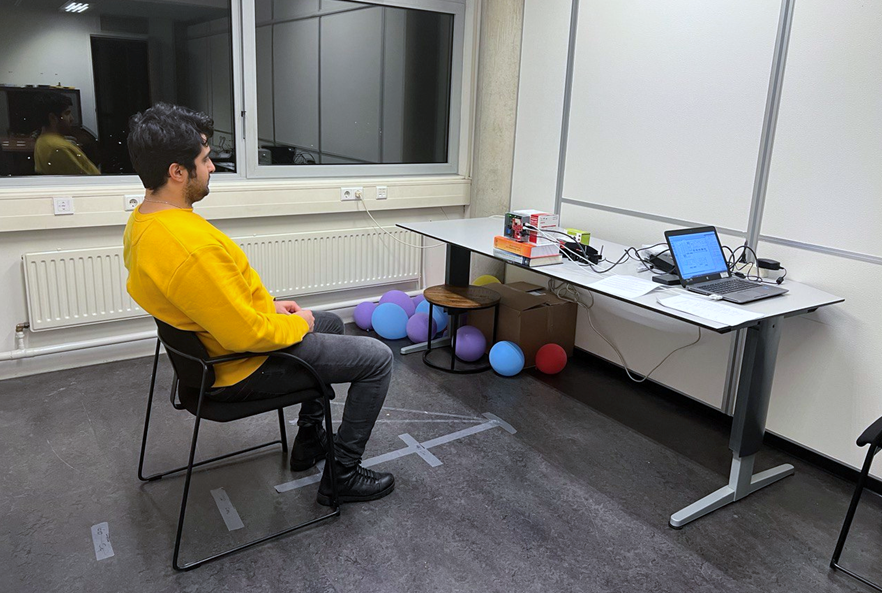}     
\caption{Participant in the Distance scenario, positioned 120 cm from the radar.}
\label{par}
\end{figure}

Our dataset also accounts for special cases and physiological conditions. One participant practiced meditation (participant 2), which significantly lowers respiration rate, while two others had asthmatic conditions (participants 5 and 6), typically associated with increased respiration rates. An overview of the different dimensions of our setup and the scenarios is illustrated in Fig. \ref{map}. Additionally, Fig. \ref{par} depicts a participant in the distance scenario seated 120 cm away from the radar. By designing these experiments and considering various extreme cases and health disorders, we aimed to evaluate the FMCW radar and our novel methods across diverse scenarios and extreme potential cases, thus highlighting the challenges that may arise when using this setup in real-world scenarios.

\section{Results}

To refine the performance of the proposed signal processing method, we conducted a comprehensive grid search on two random participants, testing approximately one million different parameter settings. Each setting involved a variation of parameters to identify the optimal configuration.
Although not all parameters could be detailed due to their vast number, the below findings are crucial in understanding the signal processing methods' behavior in diverse conditions.

Subsequent to the range FFT, we analyzed the impact of averaging different numbers of chirps for the average chirp method (16, 32, 64, 128). Our results indicate that averaging 32 chirps strikes the optimal balance for SNR improvement. When examining phase extraction techniques, both EDACM and phase unwrapping demonstrated comparable effectiveness. In the realm of impulsive noise mitigation, the EWMA and Median filters showed superior performance over the MA and WMA methods.

We also explored the optimal point for combining the outputs of receiver channels to mitigate RX channel data confusion. This was explored at different stages: immediately following range map creation, prior to filtering, and post-heart and respiration rate estimation. The findings suggest that the process flow presented in our proposed diagram—implementing RX channel data fusion before filtering—yields the best results.

When comparing filters, we observed that elliptic and Butterworth filters performed similarly. Taking into account the normal ranges for human heart and respiration rates, we set the passband frequencies for heart rate and respiration rate filtering at (0.6, 4) Hz and (0.05, 0.7) Hz, respectively. In the subsequent sections, we will delve into the results for each scenario and discuss our various findings.
In the following sections, we will use Mean Absolute Error (MAE) and Root Mean Square Error (RMSE) to assess the accuracy of each method in every scenario. These metrics are mathematically defined as:

\begin{equation}
    \text{MAE} = \frac{1}{n} \sum_{i=1}^n |y_i - \hat{y}_i|,
\end{equation}


\begin{equation}
    \text{RMSE} = \sqrt{\frac{1}{n} \sum_{i=1}^n (y_i - \hat{y}_i)^2},
\end{equation}
where \( y_i \) represents the true values, \( \hat{y}_i \) represents the predicted values, and \( n \) is the number of observations.

\subsection{Distance Scenario}
\begin{table*}[t]
    \centering
    \begin{tabular}{|c|c|c|c|c|c|c|c|c|c|c|}
    \hline
    \multirow{2}{*}{Methods} & \multicolumn{2}{c|}{40 cm}&\multicolumn{2}{c|}{80 cm}&\multicolumn{2}{c|}{120 cm}&\multicolumn{2}{c|}{160 cm}&\multicolumn{2}{c|}{Total}\\
    \cline{2-11}
    & MAE  &RMSE  &MAE  &RMSE&MAE  &RMSE  & MAE  &RMSE  & MAE  &RMSE  \\
    \hline
         FFT&  0.97& 1.25 &0.95  &1.14  & 1.27& 1.96 &1.06 &1.62  & 1.07&  1.53     \\
         \hline
         MUSIC& 1.03 & 1.27  & 1.14 &1.35  & 0.88& 1.23 &0.96 &1.24  & 1.01&  1.28  \\
         \hline
         Prony&  0.87 & 1.07& 0.77  & 0.99 & 0.79 & 0.94 &0.79 &  1.02&0.8 & 1.01  \\
         \hline
    \end{tabular}
    \caption{Distance scenario- Overall Accuracy of Respiration Rate Estimation Methods}
    \label{MAE-dis-RR}
\end{table*}

\begin{table*}[t]
    \centering
    \begin{tabular}{|c|c|c|c|c|c|c|c|c|c|c|}
    \hline
    \multirow{2}{*}{Methods} & \multicolumn{2}{c|}{40 cm}&\multicolumn{2}{c|}{80 cm}&\multicolumn{2}{c|}{120 cm}&\multicolumn{2}{c|}{160 cm}&\multicolumn{2}{c|}{Total}\\
    \cline{2-11}
    & MAE  &RMSE  &MAE  &RMSE&MAE  &RMSE  & MAE  &RMSE  & MAE  &RMSE  \\
    \hline
    FFT& 3.08&4.99  & 7.14 & 16.76 & 5.96&7.86  &5.93 &8.27  &5.53 &10.44       \\
         \hline
    CTF-His&13.56  & 15.43 & 15.09 &16.77 &15.77 &17.55  &15.91 & 17.25 &15.08 & 16.77  \\
         \hline
    CTF-KDE& 10.32 &11.37  & 11.45 &18.24  &10.65 & 12.37 &10.64 & 13.45 & 10.77&14.11    \\
         \hline
    MUSIC& 1.43 &2.04 &1.48 & 2.45 & 1.95&2.76  &2.35 &3.67  & 1.8&2.79    \\
         \hline
    Prony& 0.61& 0.77 & 0.57 & 0.84 &0.91 & 1.16 &1.15 & 1.63 & 0.81&1.15   \\
         \hline
    \end{tabular}
    \caption{Distance scenario- Overall Accuracy of Heart Rate Estimation Methods}
    \label{MAE-dis-HR}
\end{table*}

In the distance scenario, all 10 participants were seated motionless in front of the radar at varying distances of 40 cm, 80 cm, 120 cm, and 160 cm. For each distance, four recordings were made per participant; out of these, two were randomly selected for further processing. An observation period of 30 seconds, corresponding to a window of 600 frames, was used to perform three methods for respiration rate estimation and five methods for heart rate estimation. The results, along with reference values, are presented in Table \ref{dis-rr} for respiration rate estimation and Table \ref{dis-hr} for heart rate estimation, respectively.
All the heart rate estimated and reference values are in beats per minute (BPM) and all the respirations values are in respiration per minute (RPM).

To better understand the results and accuracy of the various estimation methods, the overall accuracy was assessed using MAE and RMSE across all participants, distances, and methods. The corresponding calculations are presented in Table \ref{MAE-dis-RR} for respiration rate estimation and Table \ref{MAE-dis-HR} for heart rate estimation, respectively.

\begin{figure}[h]
    \centering
    \begin{subfigure}[b]{0.5\textwidth}
        \includegraphics[width=8.5 cm]{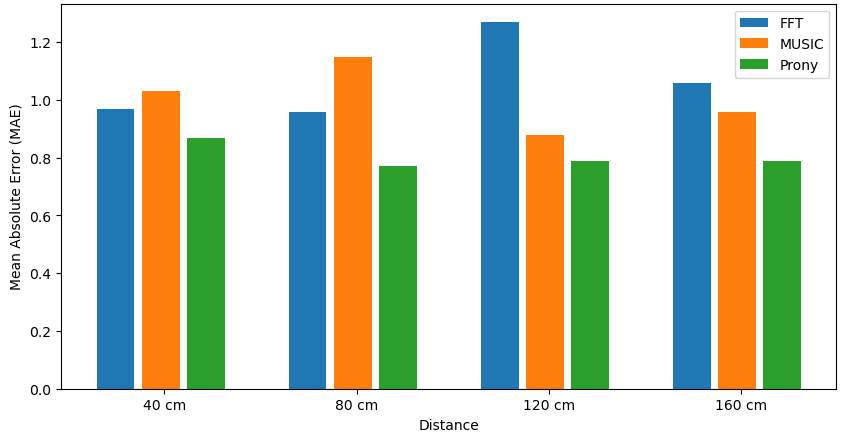}
        \caption{Respiration Rate }
    \end{subfigure}
    \hfill 
    \begin{subfigure}[b]{0.5\textwidth}
        \includegraphics[width=8.5 cm]{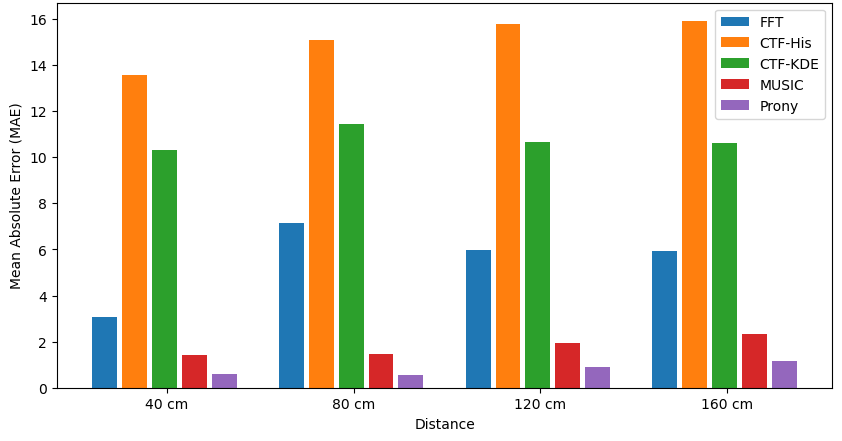}
        \caption{Heart Rate }
    \end{subfigure}
    \caption{Respiration and Heart Rate Accuracy by Method and Distance.}
    \label{dis-fig}
\end{figure}

To better illustrate the trends and changes, the results are depicted in Figure \ref{dis-fig}.
For respiration rate estimation, as anticipated, most algorithms perform well due to the dominant movement in chest cardiopulmonary displacements. Notably, the designed MUSIC and Prony algorithms show improved overall accuracy across different distances.

For heart rate estimation, although the CTF-KDE method shows improvements over the histogram-based CTF method, these still perform poorly. This underperformance compared to other studies using CTF-based methods may be attributed to the shorter observation time employed in our study, which could hinder effective performance.
The improved FFT method demonstrates significant enhancements, with an overall error rate of 5.53 BPM, a notable improvement over other FFT-based studies. 
This level of error remains too high for many applications. 
Despite this, the MUSIC and Prony methods achieve remarkably low errors of 1.01 and 0.80, respectively, representing substantial advancements compared to previous research in this field.
Figure \ref{dis-fig} also illustrates the expected trend in heart rate estimation accuracy decreasing with increased distance from the radar due to a drop in SNR. The only exception is at 40 cm, where results are worse than at 80 cm. This degradation close to the radar can be attributed to intense signal reflections and scattering, complex near-field effects, and potential signal saturation, which collectively obscure the subtle physiological variations critical for accurate measurements.

In real-world scenarios, such as hospital settings or indoor applications like patient or elderly monitoring, it is unrealistic to expect participants to remain seated facing the radar. Therefore, we also explore the effectiveness of the proposed methods in different postures, angles, and orientations in the following sections.

\subsection{Angle Scenario}

\begin{table*}[t]
    \centering
    \begin{tabular}{|c|c|c|c|c|c|c|c|c|}
    \hline
    \multirow{2}{*}{Methods} & \multicolumn{2}{c|}{0 Degree}&\multicolumn{2}{c|}{30 Degree}&\multicolumn{2}{c|}{45 Degree}&\multicolumn{2}{c|}{Total}\\
    \cline{2-9}
      &MAE  &RMSE&MAE  &RMSE  & MAE  &RMSE  & MAE  &RMSE  \\
    \hline
         FFT& 0.95&1.13  & 1.26 &2.03  & 1.02& 1.81 &1.08 &1.70         \\
        
         \hline
         MUSIC& 1.14&1.34  &  1.24&1.46  & 1.03&1.11  &1.14 &1.31      \\
         \hline
         Prony& 0.77&0.99  &0.98  & 1.60 & 0.91&1.15  & 0.88& 1.27   \\
         \hline
    \end{tabular}
    \caption{Angle Scenario- Overall Accuracy of Respiration Rate Estimation Methods}
    \label{MAE-ang-RR}
\end{table*}

\begin{table*}[t]
    \centering
    \begin{tabular}{|c|c|c|c|c|c|c|c|c|}
    \hline
    \multirow{2}{*}{Methods} & \multicolumn{2}{c|}{0 Degree}&\multicolumn{2}{c|}{30 Degree}&\multicolumn{2}{c|}{45 Degree}&\multicolumn{2}{c|}{Total}\\
    \cline{2-9}
      &MAE  &RMSE&MAE  &RMSE  & MAE  &RMSE  & MAE  &RMSE  \\
    \hline
         FFT&7.14 & 16.76 & 5.33 & 8.61 &9.58 & 17.71 &7.35 &14.93         \\
         \hline
         CTF-His& 15.09&16.77  &15.97  &17.34  &14.63 &15.76  & 15.23&16.64     \\
         \hline
         CTF-KDE& 11.45&18.24  &8.37  &9.90  & 10.77&12.83  & 10.20&14.09      \\
         \hline
         MUSIC& 1.48&2.45  & 4.44 &6.32  & 2.36&3.26  & 2.76&4.34      \\
         \hline
         Prony& 0.57& 0.84 & 0.83 &1.07  &0.90 & 1.09 &0.77 & 1.01   \\
         \hline
    \end{tabular}
    \caption{Angle Scenario- Overall Accuracy of Heart Rate Estimation Methods}
    \label{MAE-ang-HR}
\end{table*}

Considering the antenna pattern and changes in the power level of the received signal at different angles, it is crucial to evaluate how radar technology generally, and our designed methods specifically, performs when the subject is angled relative to the radar \cite{TI-DCA, dataset}. As the angle increases, we transition from the main lobe to the side lobes of the radar antenna pattern, where the received SNR is lower.

To investigate this, we established an angle scenario in which participants were positioned at angles of 0, 30, and 45 degrees, facing the wall behind the radar (not the radar antenna patch). A 30-second observation period was utilized for signal processing, and two recordings per participant were randomly selected to evaluate the performance of our designed algorithms.
The estimation results for each participant at different angles, using various methods along with reference values, are presented in Tables \ref{ang-RR} and \ref{ang-HR} for respiration rate and heart rate estimation, respectively.

In Tables \ref{MAE-ang-RR} and \ref{MAE-ang-HR}, the overall performance of each method at different angles is calculated using MAE and RMSE across all participants. Additionally, the overall accuracy for the angle scenario is computed for all angles and participants.

\begin{figure}[h]
    \centering
    \begin{subfigure}[b]{0.5\textwidth}
        \includegraphics[width=8.5 cm]{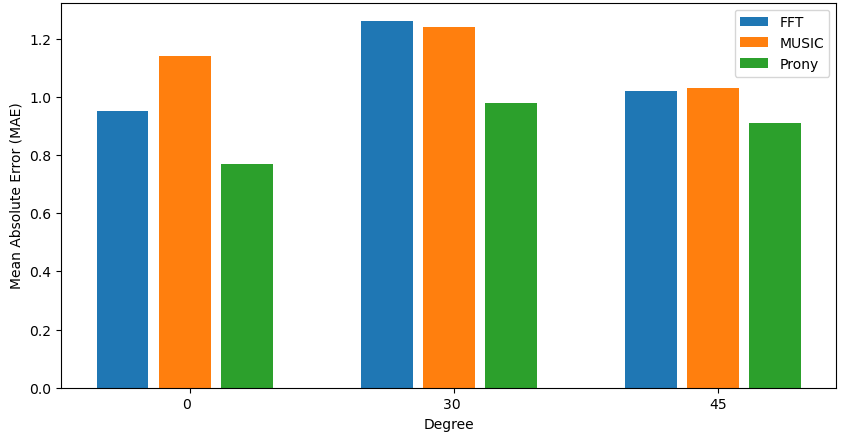}
        \caption{Respiration Rate }
    \end{subfigure}
    \hfill 
    \begin{subfigure}[b]{0.5\textwidth}
        \includegraphics[width=8.5 cm]{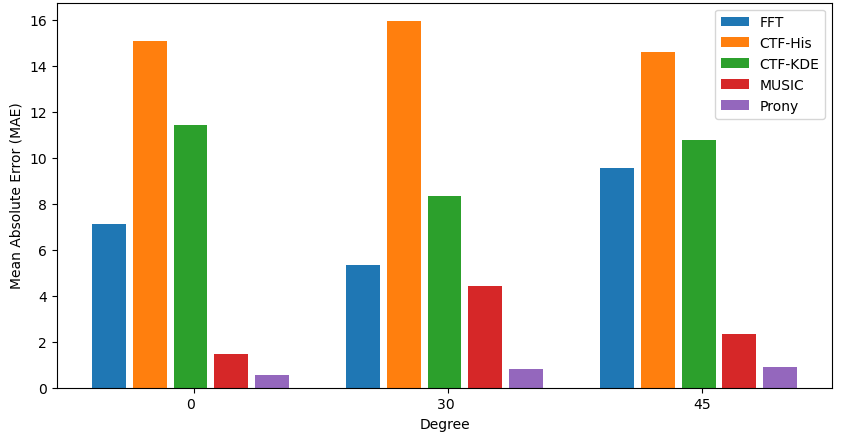}
        \caption{Heart Rate }
    \end{subfigure}
    \caption{Respiration and Heart Rate Accuracy by Method and Angle.}
    \label{ang-fig}
\end{figure}

The results, depicted in Figure \ref{ang-fig}, provide a clearer understanding of the changes. They confirm that the substantial chest displacement caused by respiration is detectable at various angles, maintaining high accuracy in these scenarios. This also demonstrates that the Prony and MUSIC methods surpass the improved FFT method, even in respiration rate estimation.
As the angle increases from 0 degrees, there is a slight drop in the accuracy of the various methods. However, the MUSIC algorithm, with an overall MAE of 2.76, and the Prony algorithm, with an overall MAE of 0.77, showcase the robustness of our proposed algorithms at different angles and SNR levels.

\subsection{Orientation Scenario}

As discussed earlier, an important real-world concern when implementing radar as a remote sensing technology in various applications is the fact that monitored subjects are not always aligned with and facing the radar.

After assessing the performance of the radar and our designed methods in the angle scenario, it is beneficial to evaluate their effectiveness in orientation scenarios. In such scenarios, which mimic real-world conditions, subjects are seated 80 cm in front of the radar, facing it with their front, left side, back, and right side sequentially. The aim is to determine whether the radar and our proposed methods can accurately capture cardiopulmonary signals even when the radar is not directly facing the chest area.
For this purpose, all participants took part in the orientation scenario under all four conditions. Again, a 30-second observation period was selected to process two randomly chosen recordings from each participant. The estimated respiration and heart rate values, along with reference values for each state, are listed in Tables \ref{or-RR} and \ref{or-HR} for all participants.

\begin{table*}[]
    \centering
    \begin{tabular}{|c|c|c|c|c|c|c|c|c|c|c|}
    \hline
    \multirow{2}{*}{Methods} & \multicolumn{2}{c|}{Front}&\multicolumn{2}{c|}{Back}&\multicolumn{2}{c|}{Left}&\multicolumn{2}{c|}{Right}&\multicolumn{2}{c|}{Total}\\
    \cline{2-11}
    & MAE  &RMSE  &MAE  &RMSE&MAE  &RMSE  & MAE  &RMSE  & MAE  &RMSE  \\
    \hline
         FFT& 0.95& 1.13 & 3.16 &4.52  & 1.37&1.83  &1.78 &2.77  &1.82 &   2.86    \\
         \hline
         MUSIC& 1.14&1.34  & 2.21 & 3.06 & 1.36&1.62  &1.47 &1.78  &1.55 & 2.06    \\
         \hline
         Prony& 0.77&0.99  & 1.26  &1.76  &1.3 & 1.76 & 0.93&1.14  &1.06 &1.45   \\
         \hline
    \end{tabular}
    \caption{Orientation Scenario- Overall Accuracy of Respiration Rate Estimation Methods}
    \label{MAE-or-RR}
\end{table*}

\begin{table*}[]
    \centering
    \begin{tabular}{|c|c|c|c|c|c|c|c|c|c|c|}
    \hline
    \multirow{2}{*}{Methods} & \multicolumn{2}{c|}{Front}&\multicolumn{2}{c|}{Back}&\multicolumn{2}{c|}{Left}&\multicolumn{2}{c|}{Right}&\multicolumn{2}{c|}{Total}\\
    \cline{2-11}
    & MAE  &RMSE  &MAE  &RMSE&MAE  &RMSE  & MAE  &RMSE  & MAE  &RMSE  \\
    \hline
         FFT& 7.14& 16.76 &6.46  & 9.79 &7.19 & 9.78 &9.70 &17.85  & 7.62&14.06       \\
         \hline
         CTF-His&15.09 & 16.77 &17.71  &19.11  &16.25 &17.45  & 16.47&17.40  & 16.38&17.70   \\
         \hline
         CTF-KDE& 11.45& 18.24 &13.65  &15.78  & 12.27&13.76  &10.65 &11.73  & 12.01&15.07    \\
         \hline
         MUSIC& 1.48& 2.45  & 1.21 & 1.54 & 1.46& 1.98 &2.34 & 3.13 &1.63 &2.35    \\
         \hline
         Prony& 0.57& 0.84 & 0.57 & 0.76 &1.28 &1.85  & 0.94&1.13  & 0.84&   1.22\\
         \hline
    \end{tabular}
    \caption{Orientation Scenario- Overall Accuracy of Heart Rate Estimation Methods}
    \label{MAE-or-HR}
\end{table*}

\begin{figure}[h]
    \centering
    \begin{subfigure}[b]{0.5\textwidth}
        \includegraphics[width=8.5 cm]{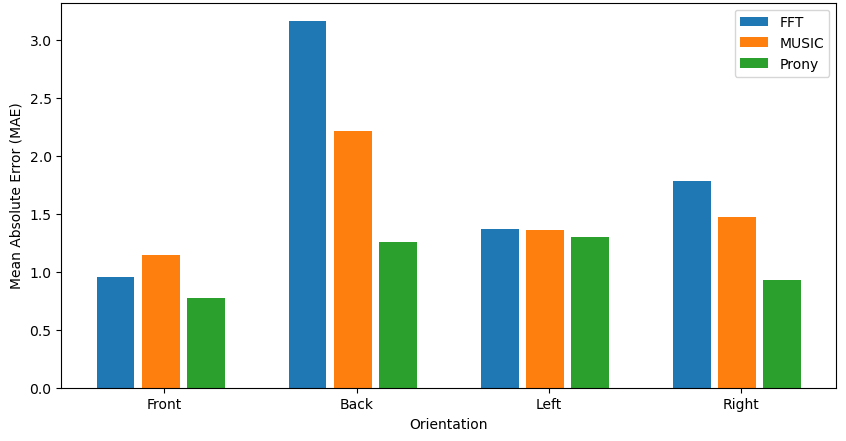}
        \caption{Respiration Rate }
    \end{subfigure}
    \hfill 
    \begin{subfigure}[b]{0.5\textwidth}
        \includegraphics[width=8.5 cm]{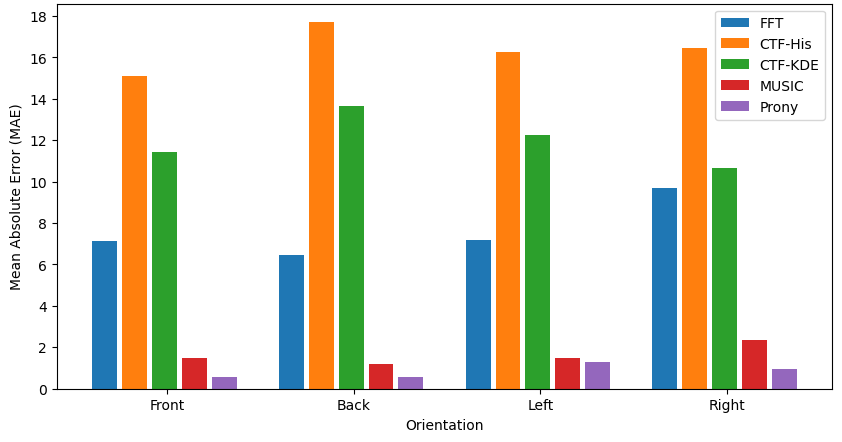}
        \caption{Heart Rate }
    \end{subfigure}
    \caption{Respiration and Heart Rate Accuracy by Method and Orientation.}
    \label{OR-fig}
\end{figure}

The MAE and RMSE for all methods across each orientation and in total for the orientation scenario, across all participants, are detailed in Table \ref{MAE-or-RR} for respiration rate and Table \ref{MAE-or-HR} for heart rate, respectively. To understand the differences and changes, these metric values are also depicted in Figure \ref{OR-fig}. The figure demonstrates that the respiration rate can still be estimated with high accuracy across different orientations. Regarding heart rate estimation, although the accuracy of the improved FFT method decreases as the subject's different sides face the radar, the designed Prony and MUSIC methods still achieve high accuracy in estimating both heart rate and respiration rate, with totals of 1.63 and 0.84 MAE for the orientation scenario, respectively.

Up to this point, we have examined the performance of various methods in different scenarios. Given that the main focus of this paper is on proposing two novel signal processing methods based on the Prony and MUSIC algorithms, and considering their demonstrated superiority over existing methods, we will now shift our attention to exploring special cases and scenarios to evaluate how these two methods perform under those conditions.

\subsection{Embracing the extremes}

Even though FMCW has demonstrated robust and accurate non-contact vital sign estimation in recent scenarios, its efficacy must also be assessed in extreme and special cases where the heart rate or respiration rate is abnormally high or low.
For this purpose, we included one participant (Participant 2) experienced in meditation and two participants (Participants 5 and 6) with asthmatic breathing disorders.

It has been observed that individuals experienced in meditation typically exhibit a lower respiration rate and higher Heart Rate Variability (HRV), which may lead to outliers and inaccuracies in estimation \cite{Med-participant}. Additionally, individuals suffering from asthma are known to have higher respiration and heart rates \cite{asthma}. This increased respiration rate can produce a different pattern of respiration rate harmonics, potentially leading to inaccurate heart rate estimation.

\subsubsection{Heart and Respiration Rate- Performance Analysis}
\begin{figure*}[t]  
    \centering
    \begin{subfigure}[b]{0.49\textwidth}  
        \includegraphics[width=\textwidth]{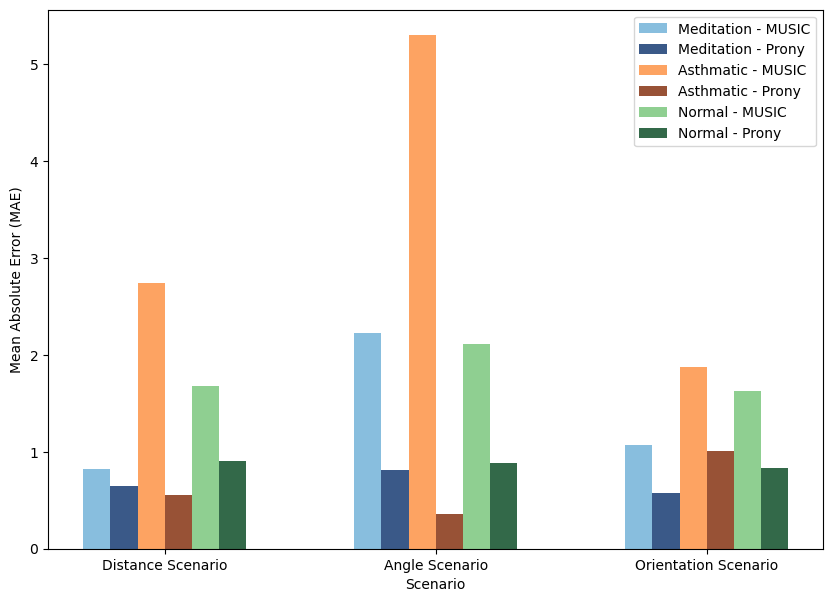}
        \caption{Heart Rate Estimation }
        \label{ext-H}
    \end{subfigure}
    \hfill 
    \begin{subfigure}[b]{0.49\textwidth}
        \includegraphics[width=\textwidth]{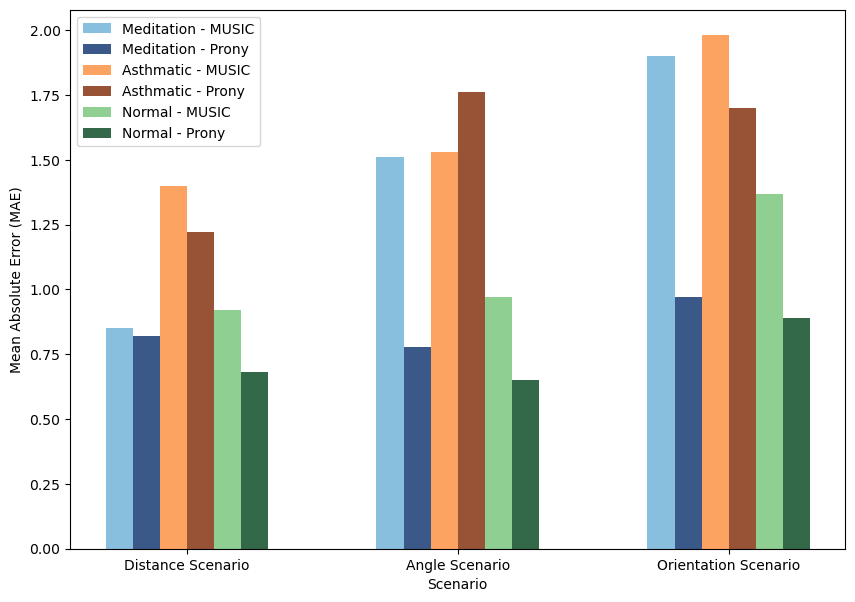}
        \caption{Respiration Rate Estimation }
        \label{ext-R}
    \end{subfigure}
    \caption{Mean Absolute Error (MAE) Across Different Participant Groups and Methods.}
    \label{ext-fig}
\end{figure*}

To evaluate the performance of the FMCW radar and our tailored algorithms (Prony and MUSIC), we divided our participants into three distinct groups. The first group comprised participants with no special conditions; the second group included participants suffering from asthma; and the third group consisted of a participant experienced in meditation. We then compared the MAE values for each group across different scenarios and methods, as illustrated in Figure \ref{ext-fig}.

For respiration rate estimation, Figure \ref{ext-R} highlights that the estimation error for the group suffering from asthma is higher compared to other groups. In the group with meditation experience, Prony continues to demonstrate robust and accurate performance, while MUSIC generally exhibits higher errors. Notably, the worst errors and the outliers still maintain an error rate of less than 2 RPM, underscoring the overall effectiveness of the methods under evaluation.

As for heart rate estimation, Figure \ref{ext-H} demonstrates consistent accuracy and performance in the estimated values among the normal participant group and the group experienced in meditation, with both groups exhibiting low error rates, though Prony typically outperforms MUSIC. This figure also reveals that the group suffering from asthma poses the greatest challenge, displaying higher error rates compared to other groups. Nevertheless, while MUSIC is notably impacted by this condition, Prony maintains robust and precise performance.

These findings indicate that both MUSIC and Prony are effective methods for heart and respiration rate estimation, consistently yielding low errors even in special cases and extreme scenarios. In heart rate estimation, the Prony algorithm consistently outperforms MUSIC, providing reliable and robust estimations.
Prony's consistent performance in respiration rate monitoring across various participant conditions highlights its potential for reliable use in diverse medical applications.
Although individual calibration may be needed for atypical patterns, such as those in asthmatic conditions, its effectiveness remains robust for heart rate estimation across all groups.

\subsubsection{Heart Rate Estimation- Level of Agreement Analysis}

\begin{figure*}[t]  
    \centering
    \begin{subfigure}[b]{0.49\textwidth}  
        \includegraphics[width=\textwidth]{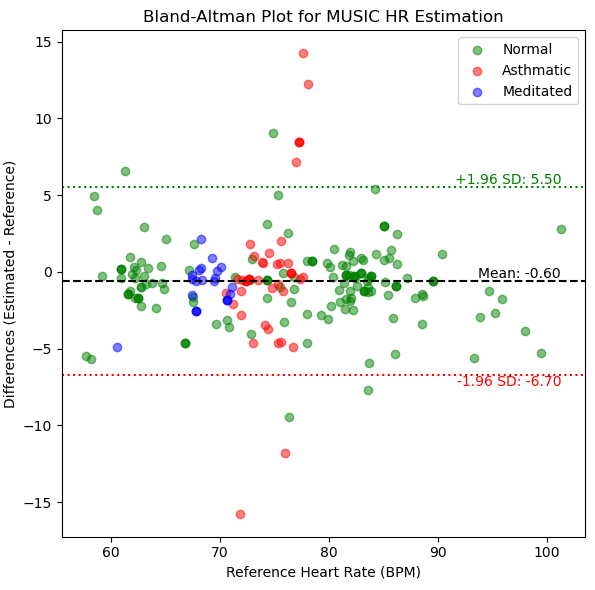}
    \end{subfigure}
    \hfill 
    \begin{subfigure}[b]{0.49\textwidth}
        \includegraphics[width=\textwidth]{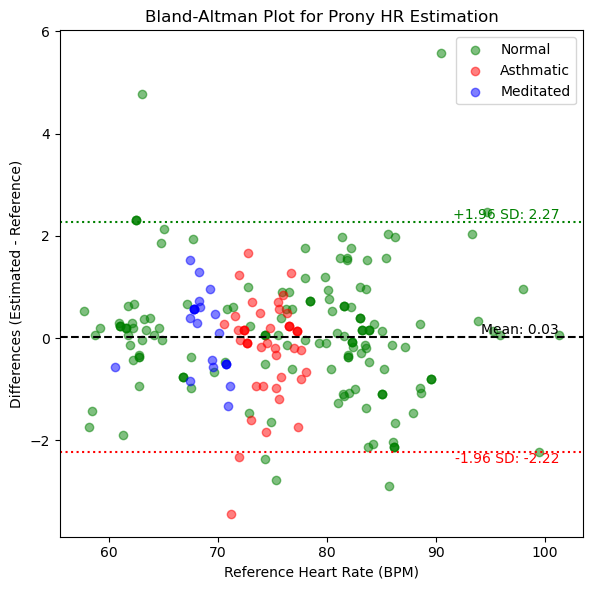}
    \end{subfigure}
    \caption{Bland-Altman plots comparing MUSIC and Prony heart rate estimation methods.}
    \label{HR-bland}
\end{figure*}

Given the complexity inherent in heart rate estimation via FMCW radar, this section explores the agreement of all measurements across various scenarios, comparing the values estimated by our developed methods, Prony and MUSIC, with those of the reference. Bland-Altman analysis is utilized as the evaluative framework to identify agreements and uncover any potential systematic or proportional biases.
Figure \ref{HR-bland} displays the Bland-Altman plots for the MUSIC and Prony algorithms, depicted separately for clarity in comparison.

The Bland-Altman plots reveal a mean bias of -0.60 bpm for the MUSIC method, implying a slight tendency to underestimate heart rate values compared to the reference. While this method exhibits a broad range of limits of agreement, from -6.70 bpm to 5.50 bpm, it is noted that there are outliers present beyond these limits. Such discrepancies indicate instances where the MUSIC method significantly diverges from the reference measurements, thus highlighting potential limitations in certain scenarios. Despite the MUSIC method's slight negative bias, the differences are relatively symmetrically distributed around the mean, without a discernible trend, suggesting homoscedasticity.

Conversely, the Prony method displays a negligible mean bias of 0.03 bpm, indicating a substantial alignment with the reference heart rate values. The tighter limits of agreement, ranging from -2.22 bpm to 2.27 bpm, alongside fewer outliers within these bounds, underscore a more consistent and reliable performance. The Prony method's differences are also symmetrically distributed, further supporting its robustness across the range of heart rates.
Comparatively, while both the Prony and MUSIC methods are deemed competent for heart rate estimation, the Prony method appears superior in terms of precision and agreement with the reference values. The narrower limits of agreement suggest that the Prony method would be preferable in applications requiring higher accuracy.

While our analysis offers valuable insights, it is crucial to acknowledge its limitations. Firstly, we did not confirm the assumed normal distribution of differences, which is fundamental to the Bland-Altman analysis. Additionally, the presence of outliers in the data from the MUSIC method necessitates further investigation, as these may signal areas where MUSIC lacks accuracy. Although the minimal bias observed in the Prony method is usually negligible, its importance escalates when precision is paramount in clinical environments. Consequently, for precise clinical monitoring, the Prony method is advisable. In contrast, the MUSIC method might be adequate for broader applications. The choice between the two methods should consider the precision needs of the specific context.

\subsection{Toward Real-time monitoring}
\begin{figure}[h]
\includegraphics[width=8.5 cm]{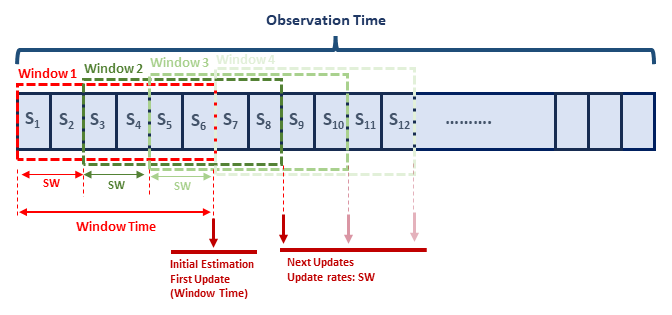}     
\caption{Visual Representation of Observational Parameters in FMCW Radar-Based Monitoring Systems}
\label{wsw-fig}
\end{figure}

FMCW radar offers a promising alternative to traditional sensing technologies due to its non-contact nature and high accuracy. Both other studies and this one highlight the functionality and significance of FMCW radar in remote vital sign monitoring. However, employing FMCW radar as a real-time monitoring system in real-world scenarios is essential for a wide range of applications, a topic that has not been extensively covered yet in previous research studies. In light of this, we investigate the real-time monitoring capabilities of FMCW radar and our designed vital sign estimation methods.

\begin{table*}[t]
\centering
    \begin{tabular}{|c|c|c|c|c|c|c|c|c|}
    \hline
    \multirow{2}{*}{Methods} & \multicolumn{2}{c|}{W=1200 (60 s)}&\multicolumn{2}{c|}{W=600 (30 s)}&\multicolumn{2}{c|}{W=400 (20 s)}&\multicolumn{2}{c|}{W=200 (10 s)}\\
    \cline{2-9}
    & MAE  &RMSE  &MAE  &RMSE&MAE  &RMSE  & MAE  &RMSE   \\
    \hline
        MUSIC & 0.2  &0.2  &1.59  &1.82&0.74&0.8  & 1.77  &2.17 \\
         Prony & 0.11  &0.11  &0.86  &0.87&1.2  &1.6  & 1.26  &1.89 \\
         \hline
    \end{tabular}
    \caption{Comparative Analysis of Heart Rate Estimation Accuracy Using Different Window Sizes.}
    \label{window-table}
\end{table*}
\begin{table*}[t]
\centering
    \begin{tabular}{|c|c|c|c|c|c|c|c|c|c|c|c|}
     \hline
    \multirow{2}{*}{Methods}&Window Size & \multicolumn{2}{c|}{SW=100 (5 s)}&\multicolumn{2}{c|}{SW=80 (4 s) }&\multicolumn{2}{c|}{SW=40 (2 s)}&\multicolumn{2}{c|}{SW=20 (1 s)}&\multicolumn{2}{c|}{SW=10 (0.5 s)}\\
    \cline{3-12}
    & W& MAE  &RMSE  &MAE  &RMSE&MAE  &RMSE  & MAE  &RMSE & MAE  &RMSE   \\
    \hline
MUSIC &600 (30 s)& 1.92  &2.29  &2.5  &2.63&2.07  &2.37  & 1.99  &2.32 & 2.02  &2.32  \\
Prony  &600 (30 s)& 0.82  &1.01  &1.08  &1.33&0.88  &1.11  & 0.89  &1.11 & 0.92  &1.13  \\
         \hline

MUSIC &400 (20 s)& 1.27  &1.47  &1.33  &1.59 & 1.5  &1.77 & 1.44  &1.7&1.45  &1.78    \\
Prony  &400 (20 s)& 1.36  &1.6  &0.87  &1.17& 0.94  &1.21 & 0.98  &1.31 &0.99  &1.3   \\
         \hline
MUSIC &200 (10 s)& 2.84  &3.04  &2.42  &2.95&2.57  &3.01  & 2.75  &3.21 & 2.63  &3.05  \\
Prony  &200 (10 s)& 1.45  &1.85  &1.47  &1.84&1.58  &2.00  & 1.4  &1.79 & 1.49  &1.88  \\
\hline
         
    \end{tabular}
    \caption{Accuracy of Heart Rate Estimations Across Different Sliding Window Factors and Window Sizes}
    \label{SW-table}
\end{table*}

\begin{figure*}[t]  
    \centering
    \begin{subfigure}[b]{0.49\textwidth}  
        \includegraphics[width=\textwidth]{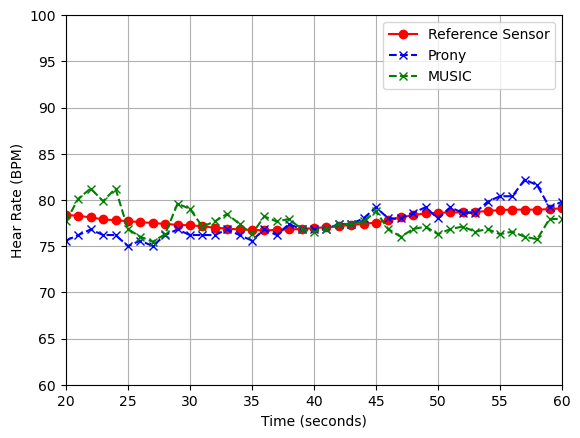}
        \caption{W=400, SW= 20}
        \label{fig:sub1}
    \end{subfigure}
    \hfill 
    \begin{subfigure}[b]{0.49\textwidth}
        \includegraphics[width=\textwidth]{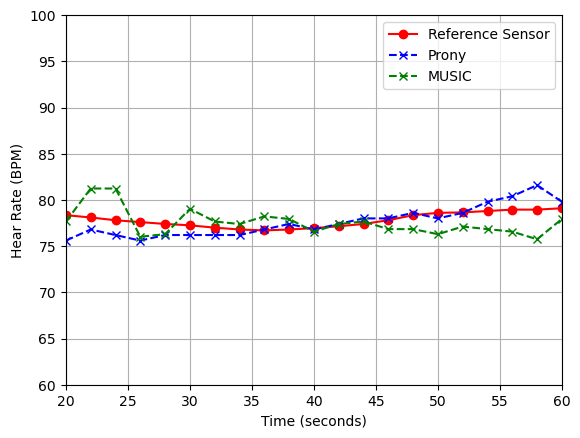}
        \caption{W=400, SW= 40}
        \label{R}
    \end{subfigure}
    \caption{Real-Time Heart Rate Estimates Using Prony and MUSIC Methods.}
    \label{RT-fig}
\end{figure*}

Real-time monitoring requires careful selection of observational parameters to balance accuracy and responsiveness effectively. This section examines two critical factors: window size and the sliding window (SW) factor, demonstrating their impact on the practical deployment of radar-based vital sign monitoring systems.
Window size refers to the number of frames the radar system processes before providing the first estimation of the vital signs. It essentially acts as the ``warm-up'' time for the radar, dictating how quickly the system can begin delivering estimations after initialization. Typically, window sizes in the literature range from 30 to 60 seconds (equivalent to 600 to 1200 frames with a frame duration of 50 milliseconds).

Sliding Window Factor (SW) determines the frequency of subsequent updates after the initial estimate. This factor defines the shift in the observation window, often a few seconds in literature, allowing for continuous and dynamic updating of the vital sign estimates, thus approaching real-time monitoring.
The concepts of the observation window, window size, and SW are visually represented in Figure \ref{wsw-fig}.

While many studies focus on SW, the importance of window size should not be overlooked. A system that only reports heart rate and respiration rate only after 60 or 30 seconds is neither practical nor beneficial, regardless of whether it updates every 5 seconds or 3 seconds.
Focusing solely on SW may not fulfill the requirements of real-time applications where timely data is essential. Therefore, this study explores the optimal configuration for real-time monitoring by considering various trade-offs between window size and sliding window factors.
For further analysis in the real-time monitoring section, we randomly selected one participant to evaluate the system's performance across different criteria. 
It is important to note that our study primarily focuses on real-time monitoring of heart rate, given its critical importance in clinical settings. Currently, we do not have access to a reference device that can evaluate respiration rate in real-time, which limits our ability to extend our findings to this vital sign at the moment.

\subsubsection{Window Size (W) }

Our methodology assesses the accuracy of heart rate estimations using window sizes of 200, 400, 600, and 1200 frames, corresponding to 10, 20, 30 and 60 seconds, respectively, and operates without a sliding window factor. This evaluation provides insights into the impact of window size on the latency of the first update and the overall system responsiveness. The results, presented in Table \ref{SW-table}, demonstrate that while a shorter window size can reduce accuracy in the estimated heart rate values, our algorithm still achieves robust performance even at the smallest window size of 200 frames. This suggests that a reduction in initial response time is possible without a significant loss in measurement accuracy.

\subsubsection{Sliding Window (SW)}

Further, we examine the impact of varying sliding window factors (10, 20, 40, 80, and 100; corresponding to shifts of 0.5 to 5 seconds) across the aforementioned window sizes. Our objective is to determine the optimal frequency of updates necessary for effective real-time monitoring while maintaining an acceptable error range. Our analysis indicates that smaller sliding windows significantly enhance the temporal resolution of updates, which is crucial for the system's ability to monitor rapid physiological changes in scenarios requiring acute medical care and continuous health monitoring.

The results for different sliding window factors and window sizes are documented in Table \ref{SW-table}. These findings confirm that even with the most favorite settings for sliding window and window size (for example, W=200 and SW=10), the Prony and MUSIC algorithms maintain a high overall accuracy.

The reference values and the estimated heart rates by the Prony and MUSIC methods are depicted in Fig. \ref{RT-fig} for a period of 40 seconds, using a window size of 400 frames (20 seconds) and sliding window factors of 20 and 40 (1 and 2 seconds respectively). As can be clearly seen in both figures, the estimated and reference values for both methods are closely aligned.
A few outliers at the beginning and end of the estimation periods are unavoidable due to suboptimal and improper window shifts at these stages.

By utilizing algorithms that perform effectively with smaller window sizes and sliding window factors within an acceptable error range, we can develop a more responsive system. Such a system is well-suited for real-world applications, offering the potential to replace traditional monitoring systems and revolutionize patient care with non-contact, continuous, real-time health data.




\begin{table*}[t]
\centering
    \begin{tabular}{|c|c|c|c|c|c|c|c|c|c|c|c|}
     \hline
    \multirow{2}{*}{Methods} &\multirow{2}{*}{Vital Signs}& \multicolumn{2}{c|}{Participant 2}&\multicolumn{2}{c|}{Participant 3}&\multicolumn{2}{c|}{Participant 4}&\multicolumn{2}{c|}{Participant 6}&\multicolumn{2}{c|}{AVG}\\
    \cline{3-12}
    & &MAE  &RMSE  &MAE  &RMSE&MAE  &RMSE  & MAE  &RMSE & MAE  &RMSE   \\
    \hline
MUSIC&Respiration Rate& 0.71 & 0.87 & 1.05& 1.29& 1.66 &1.93 &  1.43& 1.57& 1.21 &1.41 \\
Prony  &Respiration Rate&0.91 &0.97  & 0.74 &0.79& 0.31&0.43 &0.59  &0.66 & 0.63 & 0.71  \\
         \hline

         MUSIC &Heart Rate &4.74& 5.09 &2.21  &2.71&1.26 &1.54 & 6.05 &6.23 &3.56  & 3.89  \\
Prony  & Heart Rate&4.12&4.65  &1.72  & 2.16& 1.33&1.61 & 1.36 &1.75 &  2.13& 2.54\\
\hline
    \end{tabular}
    \caption{MAE and RMSE for Prony and MUSIC Methods in the Post-Exercise Elevation Scenario.}
    \label{ELV-RR-HR}
\end{table*}

\subsection{Elavated HR/ RR Scenario}

In the elevated scenario, we aim to assess whether FMCW radar and the proposed estimation algorithms are effective in the presence of high and rapidly changing vital signs. To this end, we selected four participants to engage in a combined physical activity involving running and climbing up five floors of a building. This exercise was repeated to sustain elevated vital signs, and two additional measurements were taken after each session.

The participants included one individual with asthma and another with meditation experience, to explore the algorithms' performance under varied physiological conditions. Given the rapid changes in vital signs post-exercise, a shorter window and SW were deemed necessary for accurate tracking. We opted for a window size of 400 frames (20 seconds) and an SW of 10 frames (0.5 seconds).
Signal processing and vital sign estimation were performed on one randomly chosen measurement from each participant. The MAE and RMSE values for respiration rate and heart rate estimation are detailed in Table \ref{ELV-RR-HR}.

\begin{figure}[t]
\includegraphics[width=8.5 cm]{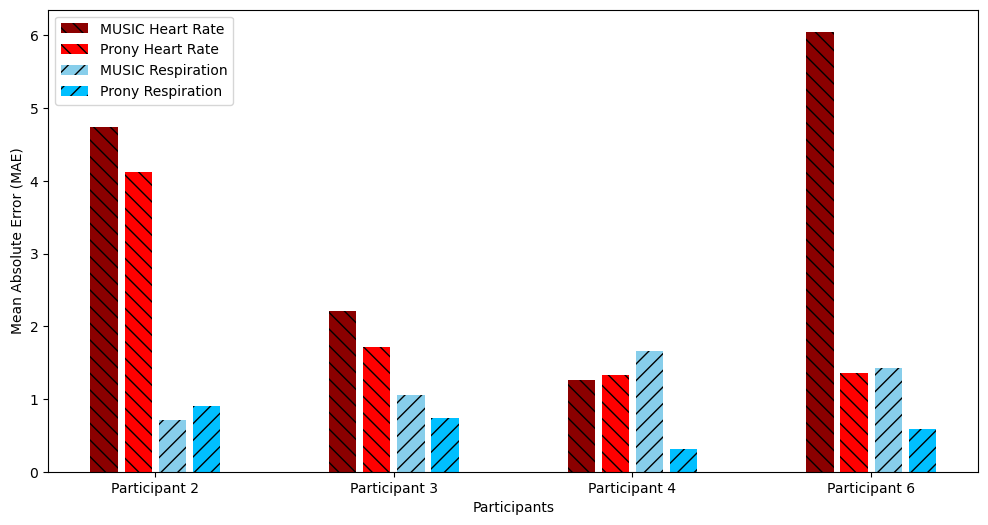}     
\caption{Participant in the Distance scenario, positioned 120 cm from the radar.}
\label{elv-fig}
\end{figure}

To assess accuracy and error in the Elevated Scenario, both heart and respiration rate estimation errors are presented in Figure \ref{elv-fig}. Although both algorithms provide stable and accurate estimations for respiration rate, heart rate estimation proves more challenging. The results demonstrate that both MUSIC and Prony successfully track and estimate the rapidly changing heart for participants 3 and 4. However, for participant 6, who suffers from asthma, the MUSIC algorithm struggles to maintain an acceptable error range, likely due to the rapid changes in signal characteristics and the increased values and amplitude of respiration rate harmonics typical of asthmatic individuals.
For participant 2, experienced in meditation, Prony and MUSIC exhibit errors of 4.74 and 4.12 BPM, respectively. In the elevation scenario post-exercise, heart rates change and drop rapidly. Given that individuals with meditation experience typically have higher HRV, this can lead to larger estimation errors.
Overall, both Prony and MUSIC, particularly Prony, effectively track and estimate post-exercise vital signs that drop rapidly.


\section{Discussion}

Our research into mm-wave FMCW radar for non-contact vital sign monitoring across diverse scenarios demonstrates substantial advancements in capturing accurate physiological data without physical contact. The adapted Prony and MUSIC algorithms were key to overcoming traditional radar limitations, handling environmental variations, and ensuring precise readings under various test conditions.
In scenarios involving different distances, orientations, and dynamic post-exercise recoveries, both algorithms effectively compensated for signal attenuation due to increased distance and rapid physiological changes, showcasing their robustness. Notably, they maintained accuracy in real-time monitoring essential for emergency medical settings where quick detection of vital sign fluctuations is critical.

Challenges related to signal interference and background noise were significant, especially in environments with varying participant conditions such as asthma or meditation practices that introduce atypical respiratory patterns. Despite these challenges, our methods effectively suppressed noise and harmonic interference, ensuring reliable measurements.
While the system performed well in standard and extreme conditions, limitations were observed in its dependency on environmental stability and outlier presence, indicating a need for further algorithm refinement. Future work will explore adaptive filtering techniques and machine learning models to enhance the system's adaptiveness to environmental changes and individual physiological differences.
This research underscores the potential of FMCW radar technology equipped with advanced signal processing for enhancing patient care through non-invasive, continuous health monitoring in various medical and healthcare settings.


\section{Conclusion}
This study demonstrates the effectiveness of using FMCW radar, enhanced with advanced signal processing techniques, for non-contact monitoring of vital signs. Extensive testing across diverse scenarios—including real-time monitoring and extreme cases involving individuals with asthma or those experienced in meditation—highlighted the superior accuracy and robustness of the proposed MUSIC and Prony methods under varied conditions. These methods not only advance the technological capabilities of vital sign monitoring but also have significant implications for their application in medical and healthcare settings where non-invasive and continuous monitoring is essential. Future efforts will focus on reducing the complexity of signal processing and improving the system's responsiveness to facilitate broader adoption in clinical environments.

\section*{Acknowledgments}
The author thanks Alex Chiumento for briefly reviewing an early draft of this manuscript.

\bibliographystyle{IEEEtran}
\bibliography{references.bib} 

\begin{table*}
\centering
\begin{tabular}{|c|c|c|c|c|c|c|c|c|c|c|}
\hline

\multirow{2}{*}{Subjects}&\multirow{2}{*}{Gender} & \multirow{2}{*}{Range (m)} & \multicolumn{2}{c|}{Reference values}  & \multicolumn{2}{c|}{Improved FFT} & \multicolumn{2}{c|}{MUSIC} & \multicolumn{2}{c|}{Prony} \\
\cline{4-11}

&& & Rec. 1& Rec. 2 &Rec. 1& Rec. 2  & Rec. 1& Rec. 2&Rec. 1& Rec. 2  \\
\hline
\hline
&& 0.4& 9&9 & 8.61&8.11 & 9.05&8.14 & 9&8.4  \\
1 &Male& 0.8& 10&9 & 8.71&8.21 & 10.36&10.36 & 9.6&9.6 \\ 
 & &1.2&  9&9 & 8.31&8.31 & 8.34&8.55 & 8.4&8.4 \\ 
 & &1.6&  9&9 & 8.31&8.91 & 8.55&9.25 & 8.4&9.6 \\ 
 \hline
 & &0.4& 6&5 & 6.11&4.90 & 5.62&5.32 & 6.6&4.2   \\
 2& Male&0.8& 4&4 & 4.70&4.40 & 5.32&5.72 & 4.8&3.0 \\ 
 & &1.2& 4.5&4.5 & 8.21&4.70 & 6.13&4.62 & 3.6&3.6 \\ 
 && 1.6& 5&4 & 4.90&9.31 & 4.62&4.92 & 3.6&4.2  \\ 
 \hline
 && 0.4& 10&8 & 9.91&7.61 & 8.55&7.84 & 10.2&7.2   \\
 3&Female& 0.8&9&8.5 & 9.41&8.21 & 8.65&7.84 & 7.8&9 \\ 
 & &1.2& 9&8 & 7.71&13.12 & 7.44&7.84 & 9.6&8.4  \\ 
 & &1.6& 8&8 & 8.41&8.01 & 7.84&7.54 & 7.2&6.6  \\ 
 \hline
 & &0.4& 14&14 & 13.82&16.82 & 13.18&12.48 & 15&12   \\
 4&Female& 0.8&14&14 & 12.52&12.62 & 12.17&12.27 & 13.8&13.2  \\ 
 & &1.2&15&15 & 13.92&14.32 & 15.09&13.68 & 14.4&15.0   \\ 
 && 1.6& 14&13.5 & 11.51&12.42 & 11.06&15.40 & 15.0&13.8  \\ 
 \hline
 & &0.4& 31&32 & 28.94&30.75 & 30.11&32.73 & 28.8&30.0   \\
 5&Female& 0.8& 30&32 & 27.84&32.55 & 28.19&33.33 & 27.6&31.8  \\ 
 && 1.2&30&32 & 29.34&36.76 & 29.20&35.75 & 28.2&31.2  \\ 
 && 1.6& 31.5&31.5 & 30.85&34.15 & 30.61&34.34 & 34.2&32.4 \\ 
 \hline
 & &0.4& 22&19 &	21.23&19.93 &		22.45&19.12	&	21.6&19.8 \\
 6&Male& 0.8&23&22&	23.13&24.24&	23.66&24.97&		22.8&24.0	
 \\ 
 & &1.2&22&23&	24.34&23.03&		21.04&21.04&		24.0&23.4 \\ 
 & &1.6&22&24 &	22.73&23.94&		21.04&25.37&		22.8&24.0 \\ 
 \hline
 & &0.4&12&12 & 9.71&14.22 & 9.35&13.68 & 12.6&13.8  \\
7 &Female& 0.8& 12&12 & 10.31&11.31 & 13.48&10.26 & 12.0&12.0  \\ 
 & &1.2& 11&11 & 10.91&11.72 & 10.26&11.27 & 12.0&9.6 \\ 
 && 1.6&10&11 & 11.92&10.31 & 10.26&9.35 & 9.6&9.0  \\ 
 \hline
 && 0.4& 13&12.5 & 12.02&11.31 & 11.47&10.46 & 13.8&11.4  \\
 8&Male& 0.8& 13&13 & 11.82&11.61 & 12.68&12.48 & 12.0&12.0  \\ 
 && 1.2& 12&11 & 11.21&11.31 & 12.07&12.27 & 11.4&11.4\\ 
 && 1.6& 12.5&10.5 & 11.92&11.21 & 13.08&10.16 & 12.0&10.2 \\ 
 \hline
 & &0.4& 6.5&5.5 & 6.61&6.31 & 6.53&6.83 & 6.0&6.0  \\
 9&Female& 0.8& 6.5&7.5 & 6.71&7.21 & 6.53&6.83 & 6.6&6.6\\ 
 && 1.2& 6.5&8.0 & 6.71&6.41 & 6.63&6.83 & 6.0&6.6 \\ 
 && 1.6& 7.0&7.5 & 7.91&7.41 & 7.44&6.93 & 7.2&6.6 \\ 
 \hline
 && 0.4& 12.5&12.0 &11.41&11.11  & 14.29&10.06  &12.6&11.4   \\
 10&Male& 0.8& 8.5&8.0& 7.51&7.11  &7.14&7.24  & 7.8&6.6  \\ 
 && 1.2& 8.5&7.5&8.81&7.21  &8.65&7.14  &8.4&6.6 \\ 
 && 1.6& 8.0&8.5 &9.51&7.91  &9.15&7.64  &7.8&7.8  \\ 
 \hline
\end{tabular}

\caption{Distance scenario- Comparison of Respiration Rate Estimations Across Different Methods}
\label{dis-rr}
\end{table*}

\begin{table*}
\centering
\begin{tabular}{|c|c|c|c|c|c|c|c|c|c|c|c|c|c|c|}
\hline

\multirow{2}{*}{Subjects} & \multirow{2}{*}{Gender}&\multirow{2}{*}{Range (m)} & \multicolumn{2}{c|}{Reference values}  & \multicolumn{2}{c|}{Improved FFT} &\multicolumn{2}{c|}{His}& \multicolumn{2}{c|}{KDE} & \multicolumn{2}{c|}{MUSIC} & \multicolumn{2}{c|}{Prony} \\
\cline{4-15}
& && Rec. 1& Rec. 2 &Rec. 1& Rec. 2  & Rec. 1& Rec. 2&Rec. 1& Rec. 2&Rec. 1& Rec. 2 &Rec. 1& Rec. 2 \\
\hline
\hline
&& 0.4& 78.03&76.5& 76.62&76.12 &76.12&68.11  & 84.74&83.84&75.31&74.54 &78.0&77.4   \\
1 &Male& 0.8&74.33&78.56 & 74.62&75.82  & 66.11&82.13 &142&65.3 & 73.78&79.16 &74.4&79.2  \\ 
 && 1.2& 75.8&79.3& 75.32&77.42  & 76.12&66.11 & 83.03&93.35&75.7&76.47 &76.2&79.2  \\ 
 & &1.6& 75.53&79.9&74.52&76.52  &82.13&54.09  & 83.13& 68.61&74.54& 76.85&75.6&79.8  \\ 
 \hline
 && 0.4&69.23& 69.56 & 85.54& 66.41 & 56.09& 52.08 & 70.51& 59.9 & 70.11& 69.15 &70.2& 69.0
 \\
 2&Male& 0.8& 70.7& 67.83 & 70.01& 63.70 & 54.09& 50.08 & 61.00& 49.78 & 68.83& 65.30 & 70.2& 68.4  \\ 
 & &1.2& 68.26& 68.1 & 81.63& 67.11 & 50.08& 54.09 & 58.79& 62.30 & 68.51& 68.19 & 69.0& 68.4  \\ 
 & &1.6& 67.46& 70.1 & 67.61& 76.92 & 48.08& 56.09 & 60.1& 64.40 & 67.23& 70.43 & 69.0& 70.2 \\ 
 \hline
 & &0.4& 88.66& 83.63 & 90.95& 79.03 & 70.11& 66.11 & 79.03& 71.31 & 87.10& 82.29 & 87.6&84.6   \\
 3& Female &0.8&89.6& 81.56 & 80.43& 81.03 & 64.10&68.11 & 76.62&68.71 & 89.02&81.33 & 88.8& 82.2 \\ 
 & &1.2& 86.23&85.93 & 73.72&72.62 & 68.11& 64.10 & 68.61& 64.50 & 88.70& 82.93 & 88.2& 85.8 \\ 
 & &1.6& 88.56&85.56 & 75.02& 85.14 & 66.11&68.11 & 79.93&74.02 & 85.18& 86.46 & 87.6& 87.6  \\ 
 \hline
 && 0.4& 63.43& 63.23 & 62.90& 62.10 & 48.08& 44.07 & 53.99& 47.67 & 63.70& 62.42 & 63.6& 63.6   \\
 4&Female & 0.8&66.76& 62.5 & 65.00& 58.79 & 46.07& 46.07 & 58.29& 62.10 & 62.10& 60.82 & 66.0& 64.8  \\ 
 && 1.2& 64.73& 65.06 & 61.90& 67.21 & 46.07& 46.07 & 61.70& 64.50 & 64.02& 67.23 & 66.6& 67.2  \\ 
 & &1.6& 67.66& 63.03 & 79.43& 47.98 & 50.08& 44.07 & 69.21& 60.70 & 69.47& 65.95 & 69.6& 67.8  \\ 
 \hline
 && 0.4& 76.3& 75.63 & 74.62& 74.62 & 90.15& 80.13 & 59.29& 69.21 & 76.85& 71.08 & 76.8& 76.2    \\
 5& Female &0.8&77.26& 76.56 & 68.41& 75.92 & 82.13& 84.14 & 74.22& 84.94 & 85.70& 76.47 & 77.4& 76.8  \\ 
 && 1.2& 76.73& 75.76 & 71.82& 74.22 & 72.12& 52.08 & 95.45& 81.53 & 71.85& 74.54 & 78.0& 75.0 \\ 
 && 1.6&78.06& 73.0 & 60.2& 73.62 & 60.1& 74.12 & 47.47& 91.55 & 90.32& 68.39 & 77.4& 71.4\\ 
 \hline
 & &0.4& 73.96&73.53 &	74.12&71.61 &	74.12&80.13 &	54.89&68.71	 &	74.54&73.01	&	73.8&72.6		
   \\
 6&Male& 0.8& 72.43& 72.7	& 72.82& 143.43	&	72.12& 80.13	&	79.33& 48.48	&	71.85& 72.24	&	72.6& 72.6	 \\ 
 && 1.2& 72.03& 74.8 &	67.91& 76.92 &		86.14& 76.12&		79.23& 97.36&		71.47& 73.78&		72.0& 75.0  \\ 
 & &1.6& 71.96& 75.33 &	72.42& 76.62 &		88.14& 76.12 &		83.23& 45.07 &		69.16& 74.54&		73.2& 75.0  \\ 
 \hline
 & &0.4&62.73& 59.2 & 60.1& 57.69 & 50.08&46.07 & 57.09& 46.77 & 63.38& 58.89 & 61.8&59.4   \\
7 &Female & 0.8& 60.96& 61.6 & 55.79& 74.02 & 40.06& 40.06 & 55.59& 55.29 & 61.14& 60.18 & 61.2& 61.8  \\ 
 & &1.2&62.1& 61.73 & 73.62& 61.50 & 48.08& 46.07 & 55.79& 53.68 & 62.42& 60.50 & 62.4& 61.8  \\ 
 & &1.6& 60.9& 63.03 & 60.70& 50.38 & 44.07& 36.06 & 51.88& 46.97 & 60.50& 62.74 & 61.2& 63.0 \\ 
 \hline
 & &0.4& 83.53& 85.06 & 79.73& 85.84 & 68.11& 60.1 & 73.22& 65.10 & 82.93& 85.82 & 83.4& 85.2   \\
 8&Male& 0.8& 83.83& 83.0 & 83.23& 81.13 & 56.09& 64.10 & 83.23& 77.93 & 83.57& 82.93 & 84.0& 83.4 \\ 
 & &1.2& 83.66& 81.83 & 76.32& 81.03 & 50.08& 58.09 & 68.41& 70.61 & 83.25& 82.93 & 85.2& 83.4  \\ 
 & &1.6& 75.36& 83.1 & 74.52& 78.33 & 62.10& 64.10 & 65.40& 71.01 & 80.37& 83.9 & 72.6& 82.8  \\ 
 \hline
 & &0.4& 82.2& 83.73 & 82.23& 94.85 & 76.12& 72.12 & 73.62& 77.62 & 79.73& 77.81 & 82.8& 81.6   \\
 9&Female &0.8&86.13& 85.1 & 100.96& 88.04 & 70.11& 70.11 & 89.75& 85.64 & 85.18& 88.06 & 84.0& 84.0  \\ 
 && 1.2& 84.26& 83.56 & 74.92& 98.66 & 74.12& 72.12 & 86.44& 91.25 & 89.66& 75.88 & 82.2& 82.2  \\ 
 & &1.6&86.03& 86.26 & 79.43& 85.44 & 74.12& 72.12 & 84.54& 85.44 & 80.69& 86.78 & 84.0& 84.6 \\ 
 \hline
 
 & &0.4& 85.2& 81.93 & 83.74& 76.02 & 64.10& 52.08 &71.41& 71.41 &84.54& 80.69 & 84.6& 81.6   \\
 10&Male& 0.8& 83.23& 82.26&84.34& 81.83  &66.11& 60.1  &69.31& 74.42 &81.97& 81.97&83.4& 82.2  \\ 
 && 1.2& 81.96& 82.06&69.61& 79.73  & 60.1& 64.10  &64.30& 70.81 &80.05& 80.37 &81.6&81.0 \\ 
 && 1.6& 80.4& 81.56 &69.11& 72.32  &60.1& 64.10  &78.33& 66.81 &80.05& 81.97 & 79.8& 81.6 \\ 
 \hline

\end{tabular}
\caption{Distance scenario- Comparison of Heart Rate Estimations Across Different Methods.}
\label{dis-hr}
\end{table*}

\begin{table*}
\centering
\begin{tabular}{|c|c|c|c|c|c|c|c|c|c|c|}
\hline

\multirow{2}{*}{Subjects} & \multirow{2}{*}{Gender}&\multirow{2}{*}{Range (m)} & \multicolumn{2}{c|}{Reference values}  & \multicolumn{2}{c|}{Improved FFT} & \multicolumn{2}{c|}{MUSIC} & \multicolumn{2}{c|}{Prony} \\
\cline{4-11}

&& & Rec. 1& Rec. 2 &Rec. 1& Rec. 2  & Rec. 1& Rec. 2&Rec. 1& Rec. 2  \\
\hline
\hline
&& 0& 10& 9 & 8.71& 8.21 & 10.36& 10.36 & 9.6& 9.6  \\
1 &Male& 30&8.5& 8 & 8.31& 8.31 & 8.95& 10.06 & 9& 9  \\ 
 && 45& 8& 8 & 8.71& 8.41 & 10.06& 8.95 & 9.6& 9 \\ 
 
 \hline
&& 0& 4& 4 & 4.70& 4.40 & 5.32& 5.72 & 4.8& 3.0 \\
2 &Male& 30& 3.5& 4 & 4.50& 7.91 & 5.12& 5.83 & 3.6& 3.6  \\ 
 && 45& 3.5& 3.5 & 4.70& 4.80 & 5.12& 4.42 & 4.8& 2.4  \\ 
 
 \hline
&& 0& 9& 8.5 & 9.41& 8.21 & 8.65& 7.84 & 7.8& 9  \\
3 &Female& 30& 8& 8 & 8.41& 8.21 & 7.74& 7.84 & 8.4& 8.4\\ 
 && 45& 8& 7 & 7.71& 7.91 & 6.83& 7.54 & 7.2& 7.2  \\ 
 
 \hline
 && 0& 14& 14 & 12.52& 12.62 & 12.17& 12.27 & 13.8& 13.2  \\
4 &Female& 30&16& 17 & 16.52& 17.83 & 14.89& 18.32 & 16.2& 16.8\\ 
 & &45&16& 18 & 15.82& 17.32 & 14.89& 17.61 & 15.6& 16.8  \\ 
 
 \hline
&& 0& 30& 32 & 27.84& 32.55 & 28.19& 33.33 & 27.6& 31.8   \\
5 &Female& 30&30& 33 & 23.53& 36.06 & 33.23& 33.73 & 25.8& 28.2 \\ 
 && 45& 31& 35 & 29.85& 36.36 & 31.62& 33.43 & 28.2& 36.6  \\ 
 
 \hline
&& 0& 23& 22&	23.13& 24.24&	23.66& 24.97&		22.8& 24.0	
 \\
6 &Male& 30& 23& 26 &	23.63& 24.54&	25.27& 26.78 &	22.2& 24.0  \\ 
 && 45& 21& 22&	20.63& 22.93&		19.83& 23.26&		21.0& 22.2\\ 
  
 \hline
& &0&12& 12 & 10.31& 11.31 & 13.48& 10.26 & 12.0& 12.0   \\
7 &Female& 30& 12& 12 & 12.12& 14.72 & 11.17& 13.58 & 12.0& 12.6\\ 
 && 45&13& 13 & 12.42& 13.32 & 11.97& 13.99 & 12.0& 12.6 \\ 
  
 \hline
&& 0& 13& 13 & 11.82& 11.61 & 12.68& 12.48 & 12.0& 12.0   \\
8 &Male& 30& 11.5& 11 & 11.61& 10.91 & 10.46& 9.15 & 10.2& 10.8 \\ 
 && 45& 11.5& 11 & 11.61& 11.11 & 12.58& 12.07 & 12.0& 12.0\\ 
 
 \hline
 && 0& 6.5& 7.5 & 6.71& 7.21 & 6.53& 6.83 & 6.6& 6.6  \\
9 & Female& 30&8.5& 8.5 & 8.11& 8.21 & 7.64& 7.84 & 8.4& 9.6  \\ 
 && 45& 8.5& 9.0 & 7.11& 9.71 & 7.03& 9.45 & 6.6& 7.8  \\ 
  
 \hline
&& 0& 8.5& 8.0& 7.51& 7.11  &7.14& 7.24  & 7.8& 6.6 \\
10 &Male& 30& 5.0& 7.0&6.81& 7.71  &6.93& 7.34  &6.0& 6.6  \\ 
 & &45&6.0& 6.0 & 6.41& 13.32 & 6.53& 6.63  &6.0& 6.0  \\ 
  
 \hline

\end{tabular}
\caption{Angle scenario- Comparison of Respiration Rate Estimations Across Different Methods}
\label{ang-RR}
\end{table*}

\begin{table*}
\centering
\begin{tabular}{|c|c|c|c|c|c|c|c|c|c|c|c|c|c|c|}
\hline

\multirow{2}{*}{Subjects} & \multirow{2}{*}{Gender}&\multirow{2}{*}{Range (m)} & \multicolumn{2}{c|}{Reference values }  & \multicolumn{2}{c|}{Improved FFT} &\multicolumn{2}{c|}{His}& \multicolumn{2}{c|}{KDE} & \multicolumn{2}{c|}{MUSIC} & \multicolumn{2}{c|}{Prony} \\
\cline{4-15}
&& & Rec. 1& Rec. 2 &Rec. 1& Rec. 2  & Rec. 1& Rec. 2&Rec. 1& Rec. 2&Rec. 1& Rec. 2 &Rec. 1& Rec. 2 \\
\hline
&& 0 &74.33& 78.56 & 74.62& 75.82  & 66.11& 82.13 &142& 65.3 & 73.78& 79.16 &74.4& 79.2 \\
1 &Male& 30&72.8& 76.23 & 72.42& 76.22 & 80.13& 82.23 & 79.83& 64.00 & 72.24& 78.78 & 73.8& 76.8 \\ 
 && 45& 78.03& 76.33 & 146.34& 60.80 & 94.15& 60.1 & 96.96& 73.82 & 73.39& 66.85 & 79.8& 76.2 \\ 

 \hline
&& 0& 70.7& 67.83 & 70.01& 63.70 & 54.09& 50.08 & 61.00& 49.78 & 68.83& 65.30 & 70.2& 68.4   \\
2 & Male&30& 60.56& 70.93 & 60.50& 70.41 & 54.09& 50.08 & 53.68& 53.68 & 55.69& 69.47 & 60.0& 69.6  \\ 
 && 45&68.3& 68.4 & 54.39& 57.19 & 50.08& 52.08 & 54.29& 52.38 & 70.43& 67.87 & 69.6& 69.0  \\
 \hline
 && 0& 89.6& 81.56 & 80.43& 81.03 & 64.10&68.11 & 76.62&68.71 & 89.02&81.33 & 88.8& 82.2   \\
3 &Female& 30& 87.16& 82.23 & 86.94& 77.42 & 66.11& 60.1 & 84.24& 71.21 & 86.78& 82.93 & 87.0& 84.0  \\ 
 && 45& 85.43& 84.33 & 84.34& 81.93 & 66.11& 70.11 & 74.62& 71.01 & 83.9& 85.50 & 87.0& 84.6  \\
 \hline
 && 0& 66.76& 62.5 & 65.00& 58.79 & 46.07& 46.07 & 58.29& 62.10 & 62.10& 60.82 & 66.0& 64.8  \\
4 &Female& 30& 57.66& 58.43 & 75.72& 57.89 & 46.07& 48.08 & 63.80& 60.90 & 52.16& 63.38 & 58.2& 57.0  \\ 
 && 45& 58.13& 67.56 & 59.09& 55.49 & 50.08& 46.07 & 57.79& 54.89 & 52.48& 65.95 & 56.4& 67.2  \\
 \hline
& &0&   77.26& 76.56 & 68.41& 75.92 & 82.13&84.14 & 74.22& 84.94 & 85.70& 76.47 & 77.4& 76.8  \\
5 & Female&30& 77.63& 75.96 & 69.91& 68.51 & 62.10& 94.15 & 94.65& 74.02 & 91.86& 64.16 & 77.4& 76.8 \\ 
 && 45& 75.5& 75.2 & 66.81& 56.39 & 60.1& 72.12 & 67.31& 61.70 & 76.08& 75.70 & 76.2& 75.0  \\
 \hline
&& 0&72.43& 72.7	& 72.82& 143.43	&	72.12& 80.134	&	79.33& 48.481	&	71.85& 72.24	&	72.6& 72.6  \\
6 &Male& 30& 71.83& 77.0 &	59.49& 56.09 &		94.15& 64.10 &		70.71& 82.73 &		56.08& 84.16 &		72.0& 76.8  \\ 
 && 45& 73.9& 74.13 &	73.42& 57.49 &		92.15& 70.11 &	102.37& 50.68 &		74.54& 70.70 &		74.4& 73.2  \\ 
 \hline
&& 0& 60.96& 61.6 & 55.79& 74.02 & 40.06& 40.06 & 55.59& 55.29 & 61.14& 60.18 & 61.2& 61.8  \\
7 &Female& 30&61.93& 62.33 & 61.00& 61.20 & 38.06& 44.07 & 54.19& 47.98 & 61.78& 62.42 & 61.8& 63.0  \\ 
 && 45&62.76& 61.76 & 62.00&75.02 & 50.08& 48.08 & 53.79& 59.39 & 61.78& 62.74 & 62.4& 62.4  \\
 \hline
&& 0& 83.83& 83.0 & 83.23& 81.13 & 56.09& 64.10 & 83.23& 77.93 & 83.57& 82.93 & 84.0& 83.4  \\
8 &Male& 30&69.66& 70.83 & 70.81& 71.92 & 60.1& 62.10 & 65.50& 68.71 & 66.27& 67.23 & 69.0& 71.4\\ 
 && 45& 70.66& 71.4 & 70.41& 70.61 & 60.1& 66.11 & 60.70& 66.01& 67.55& 71.07 & 70.2& 72.0 \\
 \hline

&& 0& 86.13& 85.1 & 100.96& 88.04 & 70.11& 70.11 & 89.75& 85.64 & 85.18& 88.06 & 84.0& 84.0   \\
9 &Female& 30& 99.43& 95.26 & 81.23& 94.15 & 70.11& 72.12 & 85.84& 79.13 & 94.15& 92.55 & 97.2& 95.4 \\ 
 && 45& 98.03& 94.73 & 99.76& 92.75 & 74.12& 74.12 & 90.35& 85.34 & 94.15& 93.51 & 99.0& 97.2 \\
 \hline
&& 0&  83.23& 82.26&84.34& 81.83  &66.11& 60.1  &69.31& 74.42 &81.97& 81.97&83.4& 82.2  \\
10 &Male& 30& 79.8& 74.36&70.11& 74.82  &58.09& 64.10 &67.31& 69.11 & 80.37& 77.48 &81.0& 72.0  \\ 
 && 45& 82.6& 72.86&79.93& 72.62  &60.1& 60.1  &79.93& 66.11& 81.65& 68.83 &81.6& 71.4  \\
 \hline

\end{tabular}
\caption{Angle scenario- Comparison of Heart Rate Estimations Across Different Methods}
\label{ang-HR}
\end{table*}

\begin{table*}
\centering
\begin{tabular}{|c|c|c|c|c|c|c|c|c|c|c|}
\hline

\multirow{2}{*}{Subjects} &\multirow{2}{*}{Gender}& \multirow{2}{*}{Range (m)} & \multicolumn{2}{c|}{Reference values }  & \multicolumn{2}{c|}{Improved FFT} & \multicolumn{2}{c|}{MUSIC} & \multicolumn{2}{c|}{Prony} \\
\cline{4-11}

& && Rec. 1& Rec. 2 &Rec. 1& Rec. 2  & Rec. 1& Rec. 2&Rec. 1& Rec. 2  \\
\hline
\hline
& &Front&10& 9 & 8.71& 8.21 & 10.36& 10.36 & 9.6& 9.6  \\
1 &Male& Back& 8& 8.5 &20.73& 14.22 & 17.51& 14.19 & 13.2& 12 \\ 
 & &Right& 8.5& 8.5 & 15.02& 8.01 & 11.06& 10.06 & 10.2& 9.6 \\ 
 & &Left&   8.5& 8 & 8.21& 7.91 & 10.96& 10.96 & 10.2& 10.2 \\ 
 \hline
&& Front&4& 4 & 4.70& 4.40 & 5.32& 5.72 & 4.8& 3.0  \\
2 &Male& Back&3& 3.5 & 8.91& 7.51 & 5.12& 6.93 & 4.2& 3.6  \\ 
 & &Right& 4& 3 & 6.61& 4.50 & 6.03& 4.92 & 2.4& 3.6  \\ 
 & &Left& 3& 3.5 & 4.30& 4.90 & 4.21& 4.92 & 4.2& 4.8  \\  
 \hline
& &Front& 9& 8.5 & 9.41& 8.21 & 8.65& 7.84 & 7.8& 9  \\
3 &Female& Back& 8& 7 & 8.01& 11.92 & 7.14& 9.15 & 7.2& 6.6  \\ 
 & &Right&8& 8 & 8.41&13.52 & 8.14& 9.35 & 7.8& 6.6 \\ 
 && Left& 9& 8 & 8.51& 11.61 & 8.34& 8.45 & 10.2& 7.8 \\ 
 \hline
& &Front& 14& 14 & 12.52& 12.62 & 12.17& 12.27 & 13.8& 13.2  \\
4 &Female& Back&14& 13 & 14.42& 13.02 & 13.08& 11.97 & 15.0& 14.4  \\ 
 && Right& 13& 13 & 16.72& 13.02 & 12.78& 12.58 & 12.0&12.0  \\ 
 && Left& 14& 15 & 12.12& 13.92 & 13.28& 13.08 & 13.2& 14.4 \\  
 \hline
& &Front& 30& 32 & 27.84& 32.55 & 28.19& 33.33 & 27.6& 31.8  \\
5 &Female&Back& 30& 30 & 27.34& 31.95 & 27.39& 31.72 & 27.0& 30.6 \\ 
 & &Right& 33& 33 & 40.06&34.15 & 37.46& 34.34 & 30.0& 32.4  \\ 
 & &Left& 35& 34 & 37.86& 38.46 & 37.36&37.46 & 38.4& 39.0  \\  
 \hline
&& Front& 23& 22&	23.13& 24.24&	23.66& 24.97&		22.8& 24.0	
   \\
6 & Male&Back& 26.5& 26.5&	27.14& 24.64&		27.79& 28.09	&	25.8& 24.6 \\ 
 & &Right& 24& 27&	23.74& 26.84&		26.68& 28.70&		24.0& 26.4  \\ 
 & &Left& 21.5& 20.5 &	19.83& 20.33&		20.13& 20.94 &		22.2& 23.4  \\  
 \hline
& &Front& 12& 12 & 10.31& 11.31 & 13.48& 10.26 & 12.0& 12.0   \\
7 &Female& Back&11& 12 & 12.02& 12.82 & 9.86& 10.56 & 10.2& 11.4 \\ 
 && Right& 11& 11 & 9.91& 11.31 & 8.85& 10.86 & 12.6& 12.0 \\ 
 && Left& 11& 11 & 10.11& 12.62 & 9.05& 12.07 & 11.4& 11.4  \\ 
 \hline
&& Front& 13& 13 & 11.82& 11.61 & 12.68& 12.48 & 12.0& 12.0  \\
8 &Male& Back& 11& 11.5 & 14.52& 12.32 & 9.45& 11.67 & 11.4& 11.4 \\ 
 && Right& 11& 11.5 & 11.82& 11.92 & 10.16& 12.98 & 11.4& 10.8  \\ 
 & &Left& 12& 11.5 & 11.01& 12.12 & 10.26& 10.96 & 11.4& 12.6\\  
 \hline
&& Front& 6.5& 7.5 & 6.71& 7.21 & 6.53& 6.83 & 6.6& 6.6   \\
9 &Female& Back& 8.0& 8.5 & 7.61& 7.21 & 7.74& 8.04 & 7.2& 9.6 \\ 
 && Right& 9.0& 9.0 & 7.41& 8.91 & 7.34& 8.55 & 9.6& 8.4 \\ 
 && Left& 7.5& 8.5 & 7.01& 8.41 & 7.03& 7.74 & 6.6& 9.0 \\ 
 \hline
&& Front& 8.5& 8.0& 7.51& 7.11  &7.14& 7.24  & 7.8& 6.6   \\
10 &Male& Back& 6.0& 5.0&14.42& 11.11  &8.65& 8.75  &6.6& 6.0  \\ 
 && Right& 5.5& 5.0 &6.61&5.91  &6.93& 5.83  & 6.0& 5.4  \\ 
 && Left& 8.0& 6.5& 11.01& 6.91  &8.65& 7.24  &7.2& 6.6  \\ 
 \hline

\end{tabular}
\caption{Orientation scenario- Comparison of Respiration Rate Estimations Across Different Methods}
\label{or-RR}
\end{table*}

\begin{table*}
\centering
\begin{tabular}{|c|c|c|c|c|c|c|c|c|c|c|c|c|c|c|}
\hline

\multirow{2}{*}{Subjects} &\multirow{2}{*}{Gender}& \multirow{2}{*}{Range (m)} & \multicolumn{2}{c|}{Reference values}  & \multicolumn{2}{c|}{Improved FFT} &\multicolumn{2}{c|}{His}& \multicolumn{2}{c|}{KDE} & \multicolumn{2}{c|}{MUSIC} & \multicolumn{2}{c|}{Prony} \\
\cline{4-15}
&& & Rec. 1& Rec. 2 &Rec. 1& Rec. 2  & Rec. 1& Rec. 2&Rec. 1& Rec. 2&Rec. 1& Rec. 2 &Rec. 1& Rec. 2 \\
\hline
& &Front&74.33& 78.56 & 74.62& 75.82  & 66.11& 82.13 &142& 65.3 & 73.78& 79.16 &74.4& 79.2   \\
1 &Male& Back&72.96& 74.3 & 71.92&74.02 & 64.1&60.1	& 67.21&59.9 & 73.78&72.62 & 73.2&73.8   \\ 
 && Right& 80.23&75.9 &150.35&73.12&	76.12&60.1	&	70.51&84.34	&	78.01&72.62	&81.0&76.8	
\\ 
 & &Left& 76.83& 76.8	&74.52& 76.02&78.13&58.09	& 74.52&83.33	&75.70& 76.47&	77.4&76.2 \\ 
 \hline
& &Front& 70.7& 67.83 & 70.01& 63.70 & 54.09& 50.08 & 61.00& 49.78 & 68.83& 65.30 & 70.2& 68.4 \\
2 & Male&Back& 67.83& 69.43 & 62.10& 68.21 & 54.09& 46.07 & 54.19 & 51.18 & 67.23& 68.83 & 68.4& 69.0  \\ 
 && Right& 67.43& 67.4 & 54.39& 64.50 & 54.09& 52.08 & 57.59& 57.79 & 65.95& 66.91 & 66.6& 67.8 \\ 
 & &Left& 69.73& 71.13 & 70.91& 72.62 & 50.83& 56.09 & 58.79& 61.10 & 69.79& 70.11 & 70.2& 70.2 \\  
 \hline
 & &Front& 89.6& 81.56 & 80.43& 81.03 & 64.10&68.11 & 76.62&68.71 & 89.02&81.33 & 88.8& 82.2  \\
3 & Female&Back& 87.86& 88.53 & 88.34& 69.51 & 68.11& 66.11 & 77.93& 69.51 & 86.14& 87.10 & 86.4& 88.8  \\ 
 & &Right& 81.86& 83.86 & 72.02& 84.14 & 66.11& 68.11 & 72.02& 67.41 & 81.65& 82.61 & 83.4& 83.4  \\ 
 & &Left& 83.6& 83.0 & 80.03& 83.84 & 66.11& 70.11 & 72.92& 77.42 & 82.61& 83.9 & 83.4& 84.0 \\  
 \hline
&& Front& 66.76& 62.5 & 65.00& 58.79 & 46.07& 46.07 & 58.29& 62.10 & 62.10& 60.82 & 66.0& 64.8   \\
4 &Female& Back& 67.13& 58.73 & 72.82& 71.51 & 50.08& 46.07 & 67.91& 61.20 & 67.23& 62.74 & 67.8& 58.8 \\ 
 & &Right& 67.56& 64.13 & 64.70& 60.60 & 46.07& 44.07 & 51.78& 52.18 & 65.62& 61.78 & 66.6& 64.2  \\ 
 && Left& 64.83& 61.3 & 63.20& 70.81 & 44.07& 48.08 & 57.69& 62.40 & 63.70& 67.87 & 64.8& 59.4 \\ 
 \hline
 & &Front& 77.26& 76.56 & 68.41& 75.92 & 82.13&84.14 & 74.22& 84.94 & 85.70& 76.47 & 77.4& 76.8   \\
5 & Female&Back& 71.56& 70.53 & 62.20& 69.71 & 68.11& 70.11 & 69.31& 72.72 & 71.08& 69.16 & 72.0& 70.8 \\ 
 && Right& 74.5& 74.43 & 73.62& 53.18 & 60.1& 56.09 & 56.89& 68.61 & 75.70& 70.70 & 74.4& 72.6  \\ 
 && Left& 73.1& 75.6 & 51.88& 55.29 & 68.11& 58.09 & 60.40& 55.29 & 74.16& 77.62 & 73.8& 74.4 \\  
 \hline
&& Front& 72.43&72.7	& 72.82&143.43	&	72.12&80.134	&	79.33&48.481	&	71.85&72.24	&	72.6&72.6  \\
6 & Male&Back& 77.33& 77.6&	65.50& 78.03&		58.09& 48.08 &	49.98& 52.98 &		76.85& 77.24&		75.6& 	76.8  \\ 
 & &Right& 72.73& 75.36&	76.92& 60.40 &		72.12& 90.15 &		53.89& 75.02&		74.54& 70.70&		74.4& 74.4  \\ 
 && Left& 71.23& 71.93&	71.11& 56.19&		66.11& 56.09&		48.78& 51.08&		69.16& 70.70&		67.8& 69.6 \\ 
 \hline
&& Front& 60.96& 61.6 & 55.79& 74.02 & 40.06& 40.06 & 55.59& 55.29 & 61.14& 60.18 & 61.2& 61.8  \\
7 &Female& Back& 62.23& 62.2 & 59.49& 62.10 & 46.07& 44.07 & 50.88& 47.67 & 60.50& 61.78 & 61.8& 62.4 \\ 
 & &Right& 63.8& 62.76 & 70.01& 66.71 & 44.07& 44.07 & 61.50& 47.57 & 63.06& 61.78 & 64.2& 62.4  \\ 
 & &Left& 64.6& 62.73 & 68.81& 55.99 & 36.06& 44.07 & 51.08& 47.27 & 64.98& 60.50 & 64.8& 62.4 \\  
 \hline
&& Front& 83.83& 83.0 & 83.23& 81.13 & 56.09& 64.10 & 83.23& 77.93 & 83.57& 82.93 & 84.0& 83.4  \\
8 &Male& Back& 81.06& 81.23 & 74.62& 66.81 & 60.1& 60.1 & 64.80& 63.80 & 79.09& 81.65 & 79.8& 82.8  \\ 
 & &Right& 74.83& 81.43 & 73.92& 72.52 & 58.09& 58.09 & 79.03& 67.91 & 83.9& 80.69 & 73.2& 83.4  \\ 
 & &Left& 80.06& 81.53 & 68.01& 76.82 & 60.1& 58.09 & 60.50& 61.80 & 80.37& 79.73 & 81.0& 80.4 \\ 
 \hline
& &Front& 86.13& 85.1 & 100.96& 88.04 & 70.11& 70.11 & 89.75& 85.64 & 85.18& 88.06 & 84.0& 84.0   \\
9 & Female&Back&95.93& 101.33 & 91.75& 72.82 & 74.12& 72.12 & 84.74& 72.82 & 94.15& 104.09 & 96.0& 101.4 \\ 
 && Right& 93.86& 93.36 & 79.13& 83.03 & 70.11& 70.11 & 85.54& 83.03 & 90.95& 87.74 & 94.2& 95.4 \\ 
 & &Left& 85.7& 90.43 & 83.63& 75.42 & 72.12& 72.12 & 77.52& 72.32 & 87.10& 91.59 & 82.8& 96.0 \\ 
 \hline
& &Front&  83.23& 82.26&84.34& 81.83  &66.11& 60.1  &69.31& 74.42 &81.97& 81.97&83.4& 82.2   \\
10 &Male& Back& 81.96& 82.36&83.84& 84.74  &62.10& 60.1 &65.81& 65.40 &83.25& 81.97 &81.6& 82.2  \\ 
 && Right& 80.46& 78.03&79.03& 77.02  &62.10& 62.10  &70.71& 62.70&81.97& 78.77 &81.0& 79.2 \\ 
 && Left& 81.5& 80.9&68.31& 73.62  &60.1& 62.10  &68.31& 73.72 &79.09& 79.73 &80.4& 81.0\\ 
 \hline

\end{tabular}
\caption{Orientation scenario- Comparison of Heart Rate Estimations Across Different Methods}
\label{or-HR}
\end{table*}

\end{document}